\documentclass[preprint]{aastex}

\shorttitle{Hard X-ray Excess in Type 1 AGN}
\shortauthors{Tatum et al.}

\begin{document}

\title{The Global Implications of the Hard X-ray Excess in Type 1 AGN}

\author{M.M. Tatum\altaffilmark{1}, T.J. Turner\altaffilmark{1}, L. Miller\altaffilmark{2}, J.N. Reeves\altaffilmark{1,3}}

\altaffiltext{1}{Department of Physics, University of Maryland Baltimore County, Baltimore, MD 21250}

\altaffiltext{2}{Dept. of Physics, University of Oxford, 
Denys Wilkinson Building, Keble Road, Oxford OX1 3RH, U.K.}

\altaffiltext{3}{Astrophysics Group, School of Physical and Geographical Sciences, Keele 
University, Keele, Staffordshire ST5 5BG, U.K}

\begin{abstract}
Recent evidence for a strong 'hard excess' of flux  at energies $\ga 20$ keV in some {\it Suzaku} observations of type 1 Active Galactic Nuclei (AGN) has motivated an exploratory study of the phenomenon in the local type 1 AGN population. We have selected all type 1 AGN in the {\it Swift} Burst Alert Telescope (BAT) 58-month catalog and cross-correlated them with the holdings of the {\it Suzaku} public archive. We find the hard excess phenomenon to be a  ubiquitous property of type 1 AGN.  Taken together, the spectral hardness and equivalent width of Fe K$\alpha$ emission are consistent with reprocessing by an ensemble of Compton-thick clouds that partially cover the continuum source. In the context of such a model, $\sim$  80\% of the sample has a hardness ratio consistent with $>$ 50\% covering of the continuum by low-ionization, Compton-thick gas. More detailed study of the  three hardest X-ray spectra  in our sample reveal a sharp Fe K absorption edge at $\sim$ 7 keV in each of them, indicating that blurred reflection is not responsible for the  very hard spectral forms. Simple considerations place the distribution of Compton-thick clouds at or within the optical broad line region. 
\end{abstract}


\section{Introduction}
The spectral shape of the cosmic X-ray background suggests that a fraction of active galactic nuclei (AGN) are heavily obscured by Compton-thick material; however, the significance of that contribution to the cosmic X-ray background is debatable \citep{Gilli:2007qy, Treister:2009qy}. Nonetheless, the extensive search for Compton-thick AGN has lead some to identify these sources via the ratio between the X-ray flux and either the mid-infrared (MIR) flux, [O{\sc iii}] line or [Ne {\sc v}] line \citep[e.g.][]{Heckman:2005rt, Gilli:2010rt,LaMassa:2011fj}, quantities assumed to be negligibly affected by obscuration, and therefore representative of the AGN bolometric flux. Other groups have identified Compton-thick sources via the hardness ratios between the medium and soft X-ray bands \citep[e.g.][]{Winter:2009lq,Brightman:2012fr}, believing this bandpass to be best for the detection of X-ray absorption.

The cosmic X-ray background is known to require AGN to have a distribution of obscuring column densities that extend into the Compton thick regime, $N_H \ga 10^{24}$ cm$^{-2}$ \citep[c.f.][]{Comastri:1995pd, Gilli:2007qy, Treister:2009qy}, and it has been assumed that the most obscured AGN have hitherto largely escaped direct detection.  However, there is now widespread evidence for partial covering of the X-ray absorber even in type I AGN that in those models have previously been assumed to be largely unobscured \citep[e.g.][]{Tanaka:2004fk,Miller:2007fe,Turner:2009ys,Turner:2011uq}. 

High resolution UV spectroscopy has revealed multi-layered, complex absorption to be a common phenomenon in AGN \citep{Crenshaw:1999th, Crenshaw:2003vn}. The detection of absorption features, such as  H- and He-like species of C, N, O, Ne, Mg, Al, Si, and S (e.g., \citealt{Kaspi:2002lr}), has shown that the signatures of complex absorption extend into the X-ray band, with warm absorbers being common in AGN \citep{Blustin:2005mz,McKernan:2007oq}. Initially, complex absorption in the soft X-ray band was detected over three decades in ionization parameter (log $\xi$ $\sim$ 0 -- 3) and three decades in column density (N$_H$ $\sim$ 10$^{20}$ -- 10$^{23}$ cm$^{-2}$, e.g. \citealt{Netzer:2003wd,Turner:2009lr}). However, the detection of deep Fe {\sc xxv} and Fe {\sc xxvi} absorption lines expanded the known range of ionization parameter and column density in the AGN population.  Observations of NGC 3783 \citep{Kaspi:2002lr, Reeves:2004qy},  NGC 1365 \citep{Risaliti:2005uq}, Mrk 766 \citep{Miller:2007fe} and NGC 3516 \citep{Turner:2008qy} were among the first to show deep K-shell absorption lines from highly-ionized species of Fe. 

Recently, \citet{Tombesi:2010yq} performed a blind search for absorption signatures in the Fe K regime of radio-quiet AGN by cross-correlating the sources detected in the {\it RXTE} All-Sky Slew Survey with the {\it XMM-Newton} archive and detected absorption from Fe {\sc xxv}, Fe {\sc xxvi}, or both, in $\sim$ 40\% of the sources studied in their sample. The depth of these absorption lines constrains this highly ionized component of gas to $N_H$$\sim$ 10$^{22}$ -- 10$^{24}$ cm$^{-2}$ along the line-of-sight.  Thus, the full set of absorption signatures observed throughout the X-ray band have shown the X-ray absorbing gas to extend over a range of  column density (N$_H$ $\sim$ 10$^{20}$ -- few $\times$ 10$^{24}$ cm$^{-2}$), ionization parameter (log $\xi$ $\sim$ 0 -- 5), and outflow velocity a few hundred  km s$^{-1}$ to 0.3\,c \citep[e.g.][]{McKernan:2007oq, Tombesi:2011lr}. 

In addition to there being a wide range of gas conditions across the AGN population, it is also evident that individual sources  require multiple zones of absorbing gas \citep[e.g.][]{Netzer:2003wd, Blustin:2007nx}. \citet{Lee:2001fk} showed that MCG-6-30-15 required the presence of at least two distinct zones of gas to explain the soft-band absorption features in the {\it Chandra} {\sc meg} grating data.  \citet{Miller:2008kx} demonstrated that one could extend the complex absorber model to include zones of a higher column density, whose variations in covering fraction could then explain the marked spectral variability observed in MCG-6-30-15, with no requirement for a contribution from blurred reflection. This multi-epoch analysis revealed MCG-6-30-15 to be typical for a general class of AGN dominated by absorption effects. Similarly, a study of NGC 3516 found that the emission and absorption features detected in the grating data required the presence of four distinguishable zones of gas to shape the spectrum \citep{Turner:2008qy}. The marked spectral variability in this source was attributed to variations in the covering fractions of one of the intermediate layers. Notably, a broad dip has been observed in the X-ray light curve of both objects \citep{McKernan:1998vn,Turner:2008qy}, taken as supporting evidence for  occultation of the X-ray continuum by absorbing gas. 

Further evidence for large columns of absorbing gas along the line-of-sight has come from {\it Suzaku}. The {\it Suzaku} observations of 1H 0419--577, PDS 456 and NGC 1365 revealed a marked excess of flux above 20 keV,  compared to that predicted from fits to data below 10 keV, dubbed a `hard excess' \citep[][respectively]{Turner:2009ys, Reeves:2009zt, Risaliti:2009lr}. In these sources, the high PIN-band flux (15--50 keV) was explained by the presence of a Compton-thick absorber covering $>$ 70\% of the continuum source, suggesting that type 1 AGN can be unobscured in the optical bandpass, but partially covered in the X-ray regime. In order to obtain the true intrinsic X-ray luminosity of these sources, one must apply significant corrections for absorption and for Compton-scattering losses \citep[e.g.][]{Turner:2009ys, Reeves:2009zt}. 

The results from 1H 0419--577, PDS 456, and NGC 1365 motivated an exploratory study of the hard excess phenomenon in the local type 1 AGN population. In this paper, we investigate this phenomenon using a BAT-selected sample of type 1 AGN. We begin in Section 2 by outlining the observations and data reduction. In section 3, we measure spectral hardness ratios and Fe\,K$\alpha$ emission line equivalent widths for the sample, and present a simple Monte Carlo radiative transfer calculation that can reproduce these measurements.. In Section 4, we present the results of some case studies. In Section 5, we discusses the implications of our analysis. We draw conclusions in Section 6.
       
 \section{The Observational Data}
The {\it Swift} BAT is a sensitive coded aperture imaging instrument with a large (1.4 steradian) field-of-view.  The energy range over which BAT can perform imaging observations is 15-150 keV, yielding positions accurate to $\sim 4 '$. The bandpass of BAT allows it to make  an unbiased survey of the X-ray sky for column densities up to $\sim$ 10$^{24}$ cm$^{-2}$. We selected all type 1 AGN, including intermediates up to type 1.9 and excluding known radio sources and LINERS, from the 58-month BAT catalog\footnote{http://heasarc.gsfc.nasa.gov/docs/swift/results/bs58mon/} \citep{Baumgartner:2010fk}. To understand the sample properties, we required simultaneous medium (2--10 keV) and hard X-ray ($>$10 keV) data. Therefore, we cross-correlated those selected sources with the {\it Suzaku} archive, including all sources available in the public domain as of 2011 December 20. Our selection criteria yielded a total sample of 43 objects and 76 observations: 24 objects are classified as type 1-1.2, 16 objects are classified as type 1.5, and 3 objects are classified as type 1.8-1.9. Sources with multiple observations were reduced and analyzed individually. Details concerning the observations and data are presented in Table~\ref{tab:table}. 

{\em Suzaku} has four X-ray telescopes, each containing a silicon CCD within its focal plane forming the X-ray Imaging Spectrometers (XIS) suite. XIS0, XIS2 and XIS3 are front-illuminated (FI), providing data over a usable range of 0.6-10.0 keV with an energy resolution   FWHM  $\sim$ 130 eV at 6.0 keV. In November 2006, a charge leak was discovered in XIS2, making XIS0 and XIS3 the only operational FI chips. XIS1 is back-illuminated. The back-illuminated configuration extends the soft band to $\sim$0.2 keV; however, it also results in a lower effective area and higher background rate in the Fe K regime, compared to the FI chips. Consequently, XIS1 is excluded from our spectral analysis. {\it Suzaku} also carries the Hard X-ray Detector (HXD) that contains a silicon PIN diode detector covering a range of 10-100 keV with a usable energy range of 15--50 keV. 

The data were reduced using HEAsoft v.6.10. The XIS cleaned event files were screened in XSELECT to exclude data  during passage through the South Atlantic Anomaly and also  excluding data starting 500 s before entry  and up to 500 s after exit. In addition, we excluded data having  an Earth elevation angle $<$ 10$^{\circ}$ and a cut-off rigidity $>$ 6 GeV.  CCDs were in 3 x 3 and 5 x 5 edit modes, with normal clocking mode. Good events were selected, having  grades 0, 2, 3, 4, and 6, while  hot and flickering pixels were removed using the SISCLEAN script. The spaced-row charge injection was utilized. XIS spectra were extracted from circular regions of 3.0' radius centered on the source, while the background was extracted from a region of the same size offset from the source and from the corners of the chip that register calibration data. 

The cleaned PIN data was reduced utilizing the ftool {\sc hxdpinxbpi}. This tool calculates good time intervals (GTIs) of the non X-ray instrumental background (NXB, using model 'D' released 2008 June 17\footnote{http://www.astro.isas.jaxa.jp/suzaku/doc/suzakumemo/suzakumemo-2007-01.pdf} ) data that overlap with the source data and extracts both the source spectrum and NXB background spectrum through that common GTI. A simulated cosmic X-ray background (CXB) spectrum is calculated, based on the CXB spectrum reported in \citet{Boldt:1987kx} and renormalized to the 35$^\prime$ $\times$ 35$^\prime$ field of view of the PIN instrument, and combined with the NXB spectrum to produce a total PIN background spectrum for the observation. A dead-time correction $\sim$ 4-5\% is  applied to the source spectrum using the ftool {\sc hxddtcor}. Finally, the PIN data have a known 1$\sigma$ systematic uncertainty of 1.3\%\footnote{http://heasarc.nasa.gov/docs/suzaku/analysis/watchout.html}, which was applied to the PIN data in {\sc grppha}.  

\section{Spectral Analysis and Sample Results}
 \label{sec:fitting}
In this paper we are primarily interested in the gross spectral properties of the sample of AGN, and in particular in the broad-band hardness ratios and emission line equivalent widths.  In order to measure these quantities, we fit some simple models to each AGN spectrum.  The aim of these models is not to obtain accurate physical parameters for the X-ray sources, but rather to get sufficiently accurate measurements of broad-band fluxes, corrected for the instrumental response, and sufficiently accurate measurements of Fe\,K$\alpha$ emission line equivalent width, to be able to characterize the gross properties of the spectra.

We fitted each observation over the 2-50 keV bandpass using {\sc XSPEC} v 12.5 . The model used was a powerlaw  modified by an {\sc xstar} grid \citep{Kallman:2001qf} that was allowed to partially-cover the continuum. Further to this, we allowed a Gaussian component at $\sim$ 6.4 keV. Finally,  a  column of gas representing the Galactic line-of-sight absorption was parameterized using  the {\sc tbabs} model \citep{Wilms:2000qy} and was  fixed to the weighted average N$_H$ in the Dickey and Lockman survey \citep{Dickey:1990uq}. The {\sc xstar} table was based on v2.2.0 of the code and had a gas density $n = 10^{10}$ cm$^{-3}$, an illuminating  photon index $\Gamma$ = 2.2 and a gas turbulent velocity 300 km s$^{-1}$. In AGN, the gas density is poorly constrained and is degenerate with the ionization parameter and assumed radial location of the gas. As {\sc xstar} requires the gas density as an input parameter,  we chose a value that lies at a point where collision effects and continuum dilution are negligible. The photon index is justified in Section 3.3.2;  the turbulent velocity was based on the approximate outflow velocities of the warm absorbers.  

During spectral fitting the normalization and photon index, $\Gamma$, of the powerlaw continuum and the column density, covering fraction and ionization parameter of the {\sc xstar} table were allowed to vary.  This model provided an adequate representation of the observed continuum, from which we could determine the flux in each band of interest in a model independent way. The mean reduced $\chi^2$ value for the sample when fitted to our model was $\sim$1.5. Ten observations had reduced $\chi^2$ $>$ 2. More complex models yielded a reduced $\chi^2$ $\sim$ 1 for these 10 observations, and the difference in the hardness ratio calculated from the simple model and the more complex model was $<$ 15\% for each observation. We concluded that our simple model fits gave an accurate measurement of the integrated flux in the bands of interest. 

During spectral fitting, the PIN part of the model was scaled by an energy-independent constant as appropriate for the XIS or HXD nominal aim point used, 1.16 or 1.18, respectively. This scaling factor accounts for the cross-calibration constant required to correctly reconcile  the XIS and PIN data based on the calibration of the Crab Nebula \citep{Maeda:2008ab}. 

We extracted the observed energy fluxes (ergs cm$^{-2}$ s$^{-1}$) for the 2-10 keV and 15-50 keV bandpasses to determine the hardness ratio, Flux$_{15-50\, keV}$/Flux$_{2-10\, keV}$, for each observation. These bandpasses were chosen as the most meaningful and practical for determination of the hardness of the X-ray spectrum associated with the active nucleus. In this field, the 2-10 keV band has become a standard bandpass over which to quote a flux, this is partly because many previous X-ray detectors covered this bandpass (e.g. the {\it EXOSAT} ME, {\it ASCA} SIS and GIS, {\it XMM} EPIC CCDs). In any case, below 2 keV there are significant contributions to the X-ray flux from extended gas, starburst regions and other structures that are not immediately associated with the nucleus. There are no reliable data available between 10-15 keV. On the high end, the upper limit was chosen based on the limited sensitivity of the PIN above 50 keV. 

To obtain the intrinsic hardness of the AGN, one should correct for the Galactic absorption. However, we found that the Galactic absorption had a negligible effect on the 2-10 keV and 15-50 keV flux measurements for the entire sample. 
\subsection{The X-ray Hardness}
\begin{figure}
\epsscale{.7}
\plotone{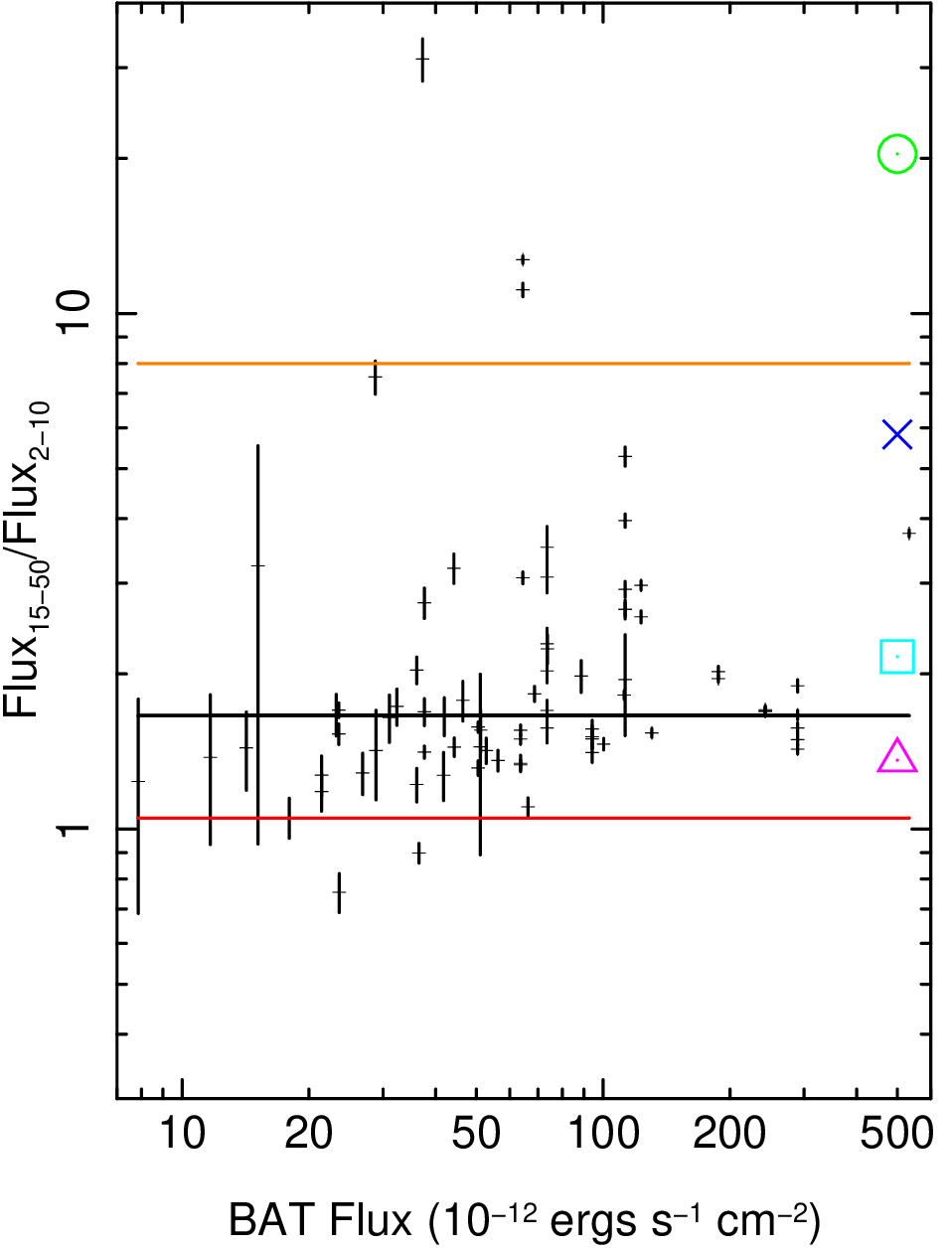}
\caption{ : The hardness ratio, Flux$_{15-50\,keV}$/Flux$_{2-10 \,keV}$, plotted against the {\it Swift} BAT flux. Superimposed is the weighted hardness ratio mean (solid horizontal line). Overlaid are model predictions for simple disk reflection for R=1 (dashed line) and R$\rightarrow \infty$ (dash-dot) and for partial covering by neutral, Compton-thick, N$_H$=2 x 10$^{24}$ cm$^{-2}$, clouds with 98\% (circle-dot), 90\% (circle-cross), 70\% (square), and 50\% (triangle) covering fractions, not accounting for Compton-scattering losses.}
\label{fig:Hardness}
\end{figure}

Figure~\ref{fig:Hardness} shows the hardness ratio, Flux$_{15-50\, keV}$/Flux$_{2-10\, keV}$, for each observation in our sample, plotted against the BAT flux \citep{Baumgartner:2010fk}. The black line is the weighted mean hardness ratio (1.65 $\pm$ 0.01) of the sample. 

To understand the distribution of hardness ratios in the sample, we compared our results to some simple models. For these models, we assumed $\Gamma$=2.0, consistent with the average photon index found for a sample of type 1 AGN \citep[cf.][]{Scott:2011fk} and a continuum cut-off at 500 keV, beyond the bandpass considered here. First we calculated the expected hardness ratio for a simple, disk reflection model.  The reflection was parameterized using {\sc pexrav} \citep{Magdziarz:1995uq}, and represents that from a standard thin disk of neutral material subtending 2$\pi$ sr at the continuum source and having Solar abundances. The disk was assumed to be observed at an inclination of 60$^{\rm o}$, as the model is insensitive to inclination (showing only a $\sim$ 10\% difference in the hardness ratio relative to our chosen value for inclinations of 30$^{\rm o}$ and 80$^{\rm o}$. The model hardness ratio shown is for the sum of the powerlaw and reflection.  In our sample, 90\% of the objects are harder than this model prediction.

We also compared our observational result to the model prediction where the continuum source is completely hidden (pure reflection), using the same parameter values as the previous disk model. Interestingly, even pure reflection cannot explain the most extreme objects in this sample (Fig.~1). We note that the sources above the pure reflection model line are typically type 1.8 and 1.9. While pure reflection may be expected from type 2 AGN, such an explanation is at odds with a sample of type 1 AGN in the context of the current paradigm. 

Of course, the hardness ratio for this model  (and the others discussed below) depends on the photon index assumed. From theory, one might expect the photon index to lie in the range 1.8$\leq$ $\Gamma$ $\leq$ 2.5 \citep[cf.][]{Haardt:1993lr}; for the simple reflection model noted above, the hardness ratio would be 1.5 for an illuminating continuum of  photon index $\Gamma= 1.8$, 1.1 for $\Gamma=2.0$, and  0.4 for $\Gamma=2.5$.

We also compared the data to predictions based upon a simple partial-covering model, characterized by a neutral absorber (pcfabs in {\sc xspec}). This simple model does not include Compton-scattering as such an effect is geometry dependent. Including Compton-scattering requires an understanding of the distribution and location of the Compton-thick gas, which currently is not well constrained for type 1 AGN. Markers on the right side of Figure~\ref{fig:Hardness} represent the ratios for the case where a neutral column of Compton-thick gas partially covers the $\Gamma=2$ continuum source, taken here to be 2 x 10$^{24}$ cm$^{-2}$. Several covering fractions are denoted: 98\% (green), 90\% (blue), 70\% (cyan) and 50\% (magenta). Approximately 80\% (35 objects, 63 observations) of the sample have hardness ratios consistent with $>$ 50\% covering of the continuum by Compton-thick gas. 

\subsection{The Equivalent Width of Fe K$\alpha$ Emission}
The broad-band X-ray spectra of local Seyfert type 1 AGN are much harder than was expected, based on an excess in the observed flux above 20 keV, compared to that predicted from fits to data below 10 keV. From a simple consideration of the distribution of hardness ratios we have demonstrated that these X-ray spectra are shaped by a high covering fraction of  neutral gas along the line-of-sight, or  are dominated by a reflected X-ray component. 

Inspection of the Fe K$\alpha$ line properties can clearly help us to distinguish the origin of spectral hardness.  For example, if we were observing the reflected component without seeing the illuminating continuum, then we would expect a large ($>$ 1 keV) line equivalent width (EW), measured against the reflected continuum \citep[e.g.][]{Krolik:1987yq,George:1991lr, LaMassa:2011fj}. In contrast, partial covering absorber models predict smaller line equivalent widths when measured against the observed continuum \cite[e.g.][]{Miller:2009cr,Yaqoob:2010fj}. 

{\it Suzaku}  has a high throughput in the Fe K regime,  allowing for accurate constraints to be placed on the Fe K$\alpha$ emission line for our sample sources. The total Fe K$\alpha$ emission line was parameterized using a Gaussian profile with a freely varying width ($\sigma$), normalization and energy. In 9 sources (11 observations), the statistical fit improved significantly ($\Delta\chi^2>$ 10 and F-test probability $< $ 0.05 ) when an additional Gaussian line component, to parameterize a broader component of  Fe K$\alpha$ emission, was allowed in the fit (with all parameters allowed to be free). In these sources, the two Gaussian components accounted for the core and broad base in the spectral shape of the total Fe K$\alpha$ emission. We determined the equivalent width of the total Fe K$\alpha$ emission by calculating it against the total observed continuum. The average total equivalent width for those observations with one Gaussian component and for those observations with narrow and broad Gaussian components were both $\sim$ 150 eV.  

Additionally, 10 observations showed significant ($\Delta\chi^2>$ 10) line emission from Fe {\sc xxv}, Fe {\sc xxvi}, or both. These features were parameterized with a narrow ($\sigma$ = 10 eV) Gaussian component fixed at 6.7 keV or 6.97 keV, respectively. We will present the results from our detailed spectral analysis, including the EW of emission in the Fe K regime, in our next paper \citep{Tatum:2012inprep}. 

Figure~\ref{fig:Spongeblob}a shows the hardness ratio plotted against the EW of the total Fe K$\alpha$ emission line. Several models are overlaid on the data, and for the models considered, the Fe K$\alpha$ emission line EW has a dependence on photon index, inclination and iron abundance of the reprocessor \citep[c.f.][]{George:1991lr, Nandra:2007lr}. As before, we assumed $\Gamma=2.0$ and solar abundances for all models.  

First, we considered the spectral hardness and line equivalent width that might be produced from simple reflection of a power-law continuum from a standard accretion disk (as described previously).  Figure~2a shows model predictions, parameterized using {\sc pexrav}, for R=1 (against the illuminating continuum) through to  pure reflection ($R \rightarrow \infty$). 
It is clear that for a given hardness ratio, disk reflection models consistently predict a much larger Fe line EW than is observed. 

We also compared our EW distribution to the values predicted by {\sc MYTorus} \citep{Murphy:2009lq}, a torodial reprocessor valid in the Compton-thick regime. In our MYTorus tables, angles in the range 0$^{\rm o} <  \theta  < 60^{\rm o}$ represent an approximately face-on view, where the continuum is not obscured by the toroidal material. As the MYTorus spectra have only a very weak dependance on inclination angle within this range  (with regard to our particular quantities for comparison), for clarity, we show only the model line for $\theta= 60^{\rm o}$ as a representation of the 'face-on' view. The small loop (Fig. 2a) shows the MYTorus predictions as the column density of the torodial gas is stepped over the range   10$^{23}$ -- 10$^{25}$ cm$^{-2}$ . For a face-on torus, the  hardness ratio peaks at $\sim  1.0$, falling below the bulk of the data points, while  line equivalent widths (18 -- 65 eV) do not achieve the span evident in the observed values. Exploring a more edge-on orientation for MYTorus, where the continuum is obscured, we fixed the inclination angle to 70$^{\rm o}$ and varied the column density of the torodial gas  as before. As illustrated in Figure 2a,  such a case does produce model predictions that span the range of values observed. Interestingly, the range of inclinations between the face-on scenario and 70$^{\rm o}$ would be consistent with the data points. However, if a large fraction of type 1 AGN require large inclination angles for the obscuring torus then this presents a challenge for the Unification Model;  the Seyfert 1 galaxies should represent the face-on cases. 

While the possibility of toroidal obscuration is intriguing, well-studied sources, such as MCG-6-30-15 \citep{Miller:2008kx}, appear to have a spectral form shaped by a complex absorber, with some zones only partially covering the continuum source.  X-ray spectral variability can then be explained as changes in the covering fraction of the absorbing gas. Such a model obviously requires the gas to be clumpy, rather than a solid and uniform form and indeed, it may be the case that there is simply a toroidal distribution of clouds in these AGN, producing observed properties close to the values for the MYTorus model.  Another possibility is that changes in the continuum source size could appear to mimic covering changes, as the fractions of obscured and unobscured light would vary. However, such a model would place tight constraints on the continuum size and torus location. 

 \subsection{A Cloud Model for the Absorbing Gas}
 \label{sec:clouds}
 
The hardness ratio, Fe K$\alpha$ line equivalent widths,  the detailed X-ray spectral shape and variability all point to a distribution of clouds, partially covering the X-ray source.  To investigate this in more detail, we performed a Monte-carlo calculation of the X-ray spectrum expected from a distribution of cold gas clouds surrounding a central source of X-rays.  This idea was first explored by \citet{Nandra:1994fj}  but computing power available at that time limited the ability to follow virtual photons as they scatter between clouds.  Now, it is possible to follow photons as they scatter any number of times between multiple clouds.
\subsubsection{The distribution of scattering clouds}
In the calculation, gas is placed in a distribution surrounding a central X-ray source.  The distribution is assumed to be a strict two-phase medium which is either vacuum or gas.  The gas phase has a constant density everywhere.  The distribution is created in two steps:
\begin{enumerate}
\item A number of points are randomly placed with a uniform spatial density within a spherical shell centered on the primary source, with a defined inner and outer radius.
\item The gas distribution is assumed to consist of spheres of constant gas density drawn around that set of points.  If two (or more) spheres overlap, the density is not doubled, but remains at the same value as elsewhere in the gas phase.  Thus in the limit where the density of sphere centers is very high, the distribution tends towards being a constant density spherical shell.
\end{enumerate}
For distributions with a high density of points, the spheres heavily overlap, and the resulting gas distribution does not then resemble a set of individual blobs, but has a sponge-like topology. For the radiative transfer calculation, the spherical distribution is assumed to be bisected by an infinite, thin accretion disk, that absorbs photons but does not itself radiate at X-ray energies, so that the scattered light received by the observer is reduced by a factor two compared with the optically-thin case without the disk. We dub the calculation ``Monte-Carlo Radiative Transfer model (MCRT)''. 

\subsubsection{The opacity, line radiation and scattering cross-section}
\label{sec:just}
The primary source of photons is assumed to be a point-like source producing a power-law X-ray spectrum with photon index $\Gamma=2.2$, consistent with the photon index found for the best studied sources once all the absorbing zones were taken into account (i.e., MCG-6-30-15 \citealt{Miller:2008kx}, Mrk 766 \citealt{Turner:2007fj}, 1H0419-577 \citealt{Turner:2009ys}). The gas is assumed to be cool, so that all ions heavier than H are neutral, but with H ionized, so that the number of free scattering electrons equals the number of H atoms.  The absorption opacity is considered to be due to all elements with Solar abundances given by \citet{Anders:1989kx}. 
In reality the Compton-scattering clouds are likely to be hot and ionized, however a full treatment of radiative transfer through such gas is computationally extremely expensive (see e.g. \citealt{Sim:2010fr}) and for the purpose of exploring a wide range of parameter space, the model presented here makes the simplifying assumption of cold gas clouds. In a future work \citep{Tatum:2012inprep}, we will compare our results with those that consider a full treatment of radiative transfer through a hot and ionized gas.

As photons propagate through the distribution they may be either absorbed by the gas, or be Compton scattered.  After absorption, there may be resulting fluorescent line emission.  This is implemented by calculating the mean free path for the photon in the gas due to three processes: 
\begin{enumerate}
\item Compton scattering, using the full energy-dependent Klein-Nishina cross-section, integrated over all scattering angles.
\item Absorption by elements other than K-shell transitions of Fe, in which case the photon disappears from the calculation (i.e. line emission from elements other than Fe K-shell transitions is neglected).
\item Absorption by an Fe K-shell transition, which may be followed by Fe\,K fluorescent line emission.
\end{enumerate}

\subsubsection{Results from MCRT}

MCRT is a representation of neutral gas clouds partially covering an X-ray continuum source. Our calculations were performed for the case of 1000 cloud centers distributed within the half a spherical shell visible to the observer. 

\begin{figure}
\scalebox{1.5}{\rotatebox{0}{\includegraphics[width=5cm]{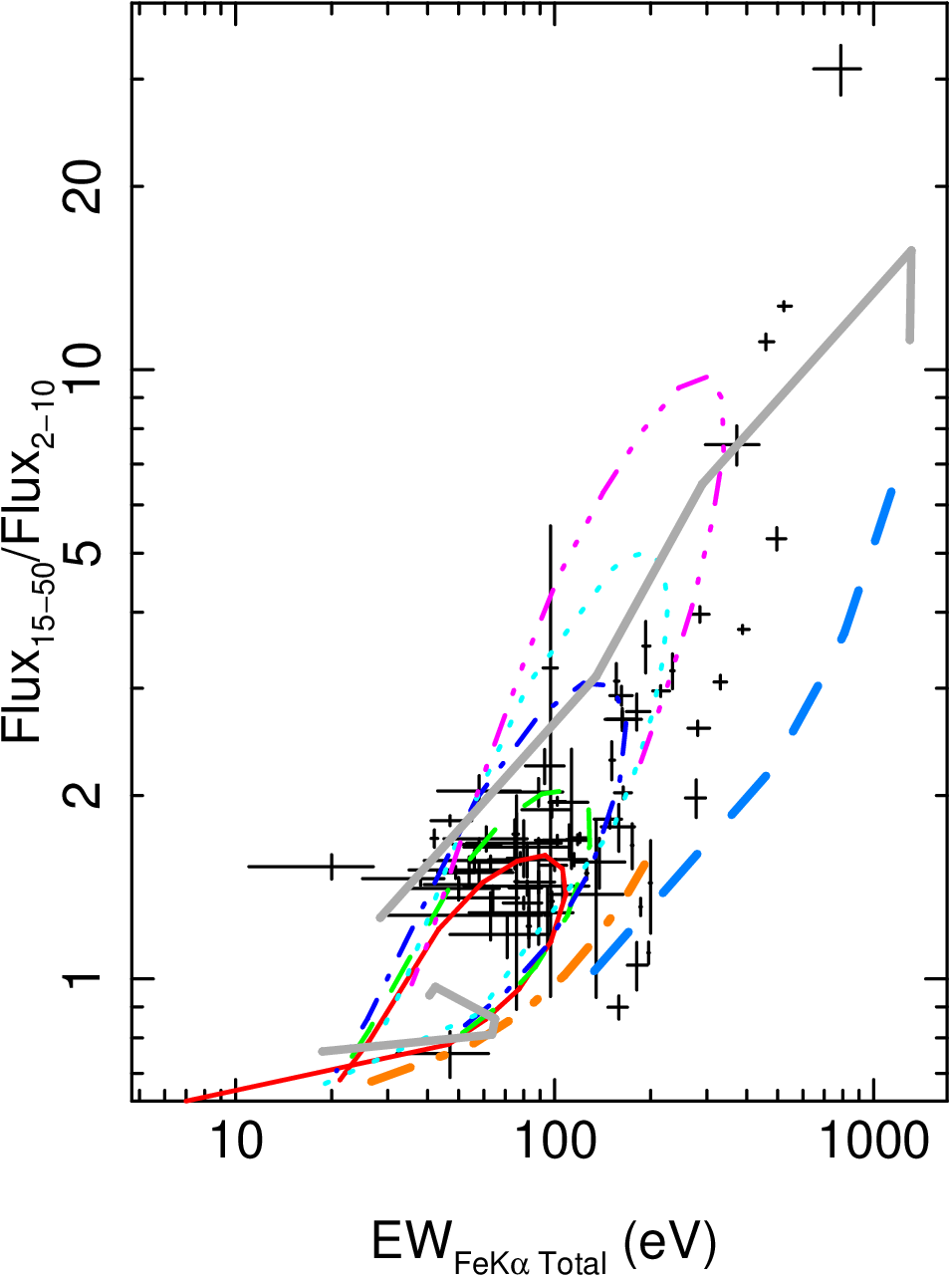}}}\,\,\,\,\,\,\,\,\,\,\,\,
\scalebox{1.57}{\rotatebox{0}{\includegraphics[width=5cm]{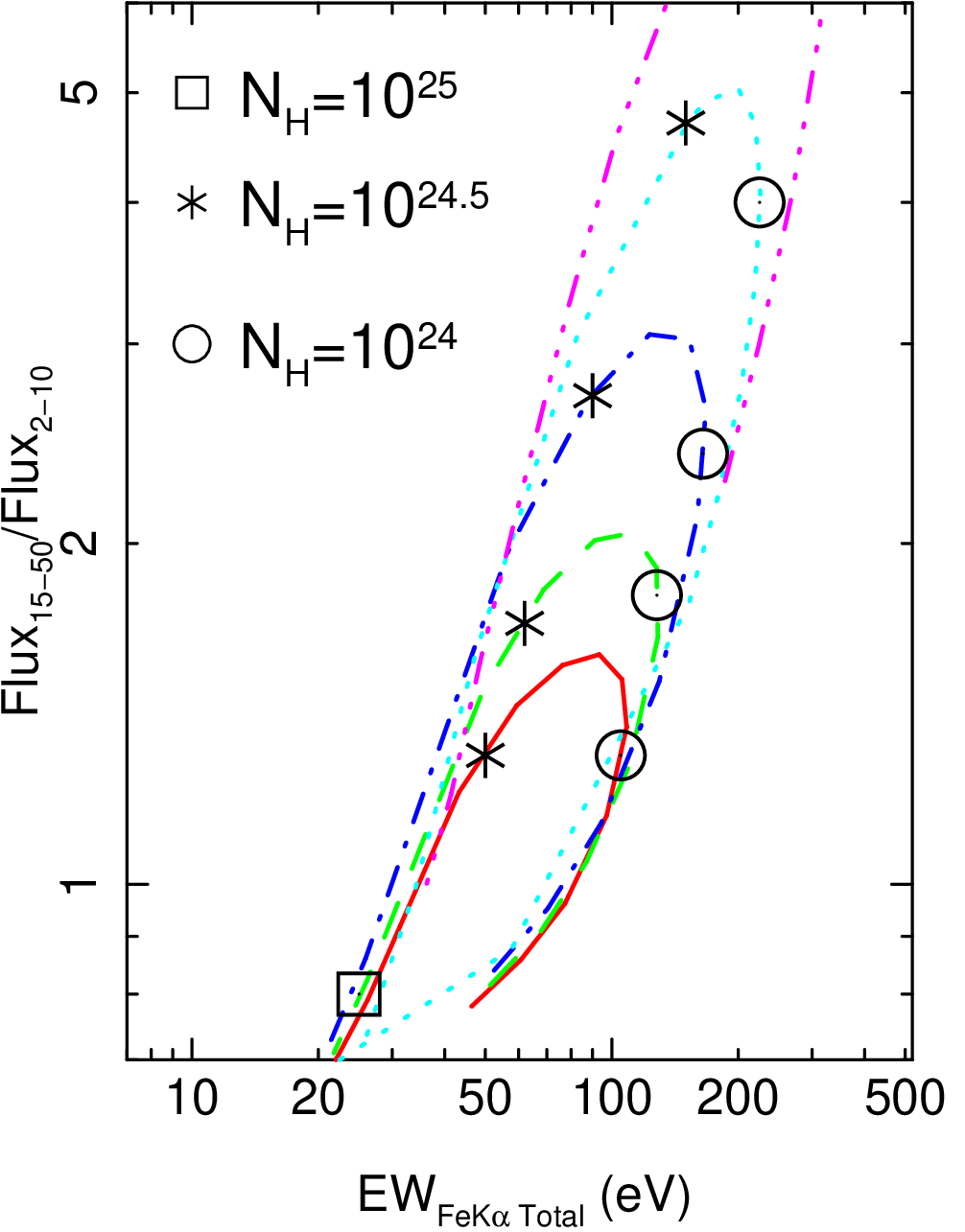}}}
\caption{ The hardness ratio plotted against the total Fe K$\alpha$ equivalent width. ({\it Left}) Model lines are overlaid for disk reflection from R=1 to pure reflection (thick dash light blue); for a Compton-thick torodial reprocessor at an inclination of 60$^{\rm o}$ (small thick solid gray) and  70$^{\rm o}$ (large thick solid gray); and for a thin absorbing shell (thick dash-dot orange). MCRT models comprise the loops (solid red, dash green, dash-dot blue, dot cyan, dash-dot-dot-dot magenta). ({\it Right}) MCRT model predictions in more detail:  each closed loop corresponds to a particular distribution of clouds.  Moving counterclockwise around any one loop corresponds to increasing column density of an individual spherical cloud. See text for details}
\label{fig:Spongeblob}
\end{figure}

The inner radius of the distribution is fixed at 10 units, scaled to the radius of an individual sphere.  The outer radii have values 12.5, 14, 15, 17.5, 20 units, from the outer loop through to the innermost loop, respectively, corresponding to leaking light fractions of  50\% -- 70\% in the 15-50 keV band. Distributions with higher volume filling factors have higher hard excesses and line equivalent widths. We calculated the hardness ratio/EW for a range of column densities covering 10$^{23}$--10$^{25}$ cm$^{-2}$ for each of the filling factors considered. The model loops connect the model predicted values for different column densities, for a given filling factor. Moving around the loop corresponds to an individual sphere having a larger column density: on the outer curve are marked sphere column densities of log$_{10}$(N$_H$/cm$^{-2}$)= 24, 24.5, 25.0. As column density increases, photoelectric absorption hardens the spectrum and increases the fluorescent line strength.  At high column densities, however, both the hard X-rays and the line flux from the high-column regions become highly attenuated, and the spectrum becomes dominated by a combination of the unobscured and scattered light, with an overall spectrum closer to the intrinsic spectrum of the source. For clarity, we plotted the model line predictions for this scenario separated from the data (Figure~\ref{fig:Spongeblob}).

The looping model lines (solid red, dash green, dash-dot blue, dot cyan, dash-dot-dot-dot magenta) in Figure~\ref{fig:Spongeblob} are model predictions from MCRT with an increasing volume filling factor (traveling from the red curve to the magenta curve). The hardness ratios and Fe K$\alpha$ emission line widths are well characterized by varying the volume filling factor of the neutral, clumpy gas, as the model lines lie within the 1$\sigma$ errors of 83\% of the data. When only accounting for the narrow core of Fe K$\alpha$ emission, the model lines would lie within the 1$\sigma$ errors of 85\% of the data, suggesting that the narrow core comprises most of the Fe K$\alpha$ emission. 

MCRT can also calculate the expected relation for a thin, uniform absorbing shell, neglecting Compton scattering. This is shown in Figure~\ref{fig:Spongeblob}, denoted as an orange model line. Moving from left to right along the curve corresponds to an increase in column density of the individual spheres ranging from 10$^{23}$--10$^{25}$ cm$^{-2}$. This model does not encompass the range of the observational data.  

Note that, in the absence of any other effects such as Doppler shifts, the results are independent of any scale size for the cloud distribution: all length scales are relative to the sphere radius, and the only dimensional parameter that enters the problem is the column density through an individual
spherical cloud. 

Consideration of different filling factors of the clumpy gas show the MCRT to be consistent with the spectral hardness and line equivalent widths of the sample sources Figure~\ref{fig:Spongeblob}.  
\section{Case Studies}
We performed more extensive spectral analysis on the hardest sources and on a typical source to investigate the nature of the Compton-thick gas in more detail.  

\subsection{A typical source: 1H 0419-577}
1H 0419-577 yielded one of the first detections of a hard excess in a type 1 AGN. This source has been observed twice by {\it Suzaku}, (2007 July 25, 2010 January 16). \citet{Turner:2009ys} found a satisfactory fit to the 2007 July 25 observation of 1H 0419-577 using  a partial covering model with 16\% and 67\% of the line-of-sight obscured with N$_H$ $\sim$ 5 x 10$^{23}$ and $\sim$ 2 x 10$^{24}$  cm$^{-2}$ and log $\xi$ $\sim$ -- 0.1 and $\sim$ 2, respectively (covering fractions not corrected for Compton-scattering losses). Analysis of the 2010 January 16 data revealed that the spectral shape remained unchanged between the two epochs, with the flux state of the source during the 2010 observation being 78\% of that observed during 2007. This result verifies the existence of the detected hard excess and establishes it as a persistent feature of 1H 0419-577. 

Examination of the sample sources showed the surprising result that the hardness ratio (1.53 $\pm$ 0.07) measured for 1H 0419-577 was consistent with the weighted mean of the sample (1.65 $\pm$ 0.01). Therefore, in the context of partial covering, the solution found for 1H 0419-577 could represent typical absorber parameters that may explain the hard excess phenomenon.

\subsection{The hardest objects}
The most extreme objects may provide the tightest constraint on the models. One naturally expects contributions from partial-covering absorption and accompanying reflection, in the context of the MCRT, and together, these give a very hard spectral form, with spectral variability expected as the absorbed fraction changes. However, the alternative suggestion that might be invoked to explain our observational result is that the hard excess is the Compton hump of relativistically blurred disk reflection, with the core of the Fe line being reprocessed light from more distant gas. \citet{Walton:2010lr} applied the blurred reflection model to 1H 0419 - 577, PDS 456 and NGC 1365 (2008 January 1). 

\begin{figure}
\scalebox{1.3}{\rotatebox{-90}{\includegraphics[width=5cm]{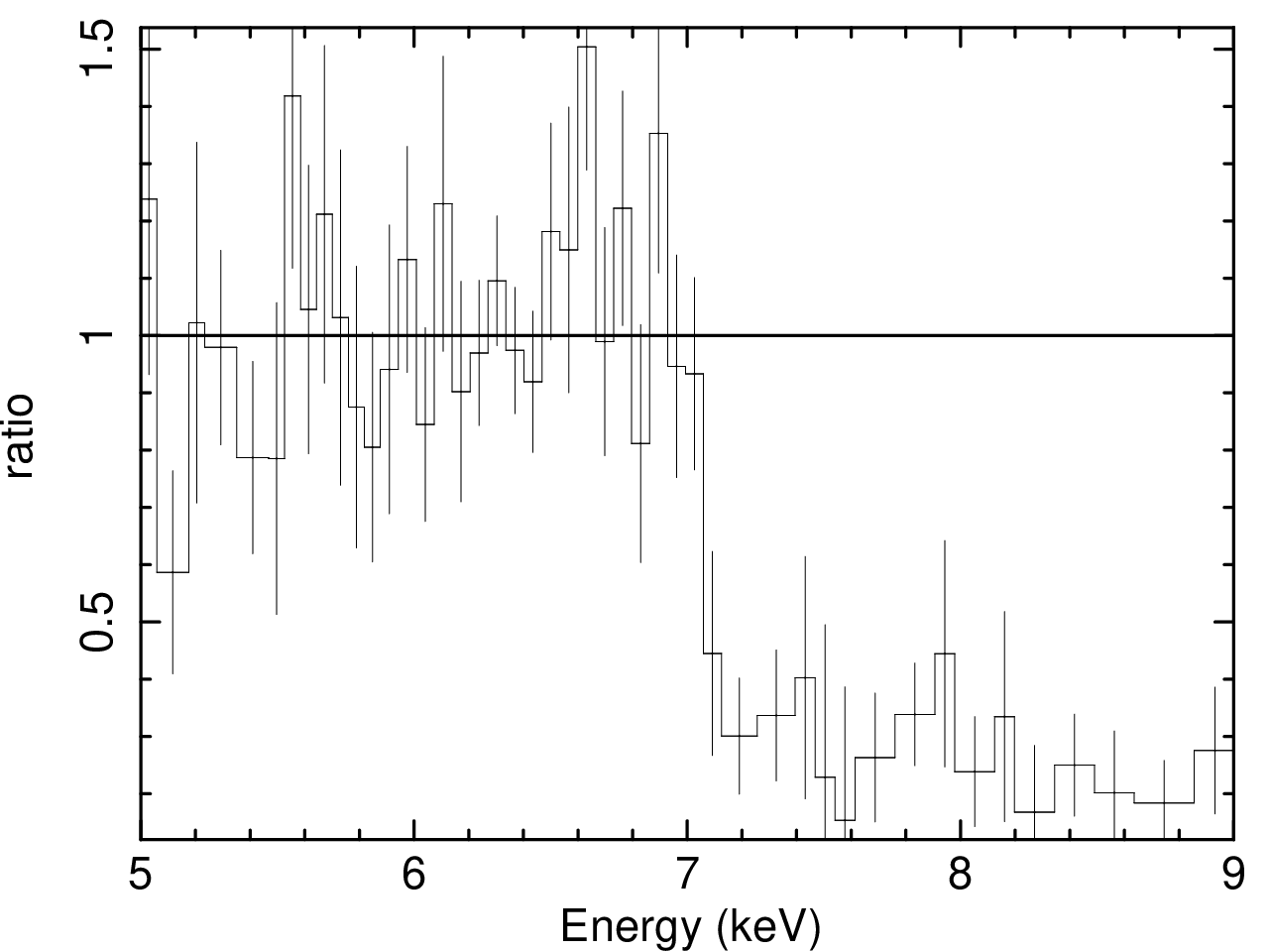}}}
\scalebox{1.3}{\rotatebox{-90}{\includegraphics[width=5cm]{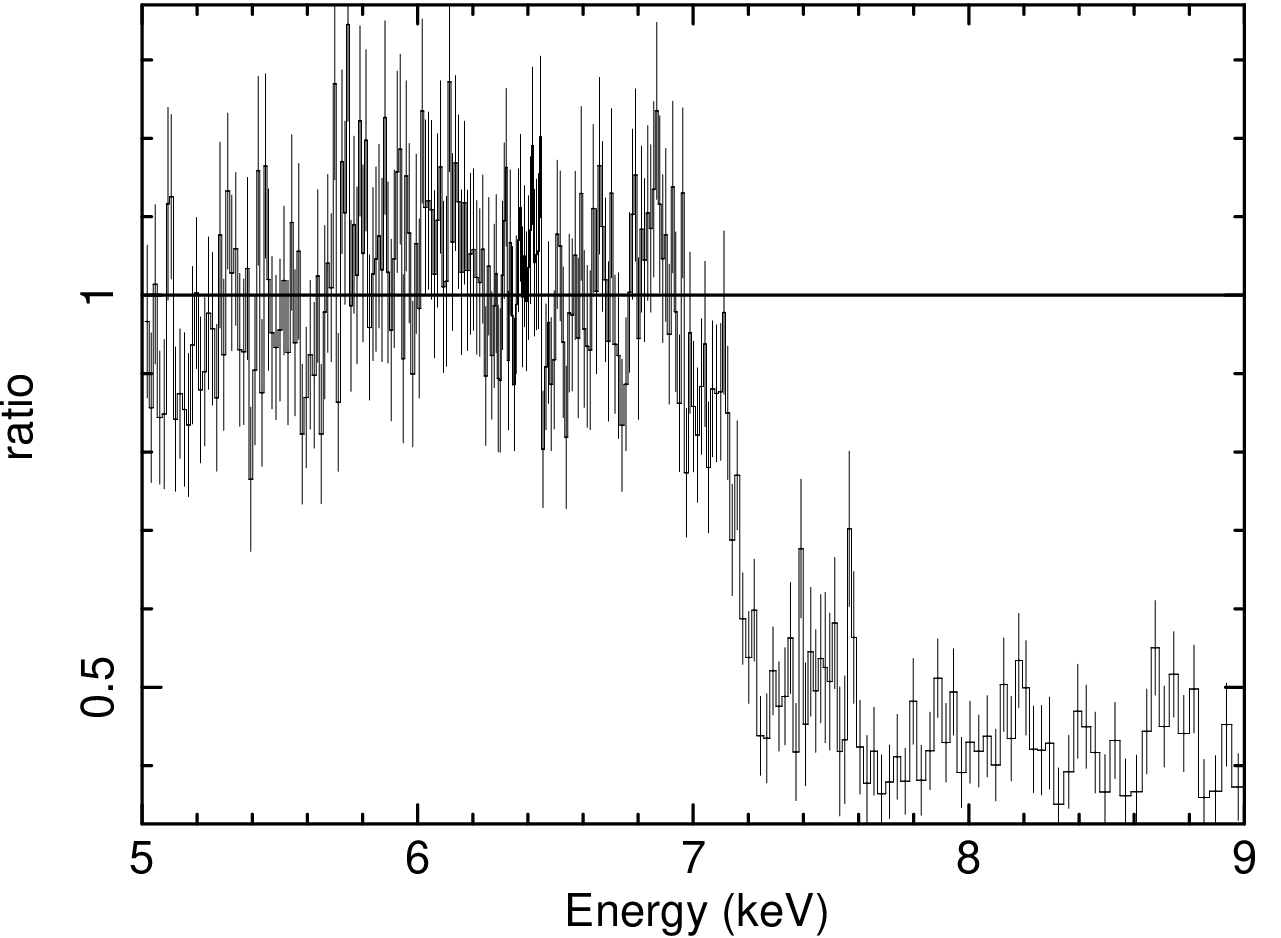}}}
\scalebox{1.3}{\rotatebox{-90}{\includegraphics[width=5cm]{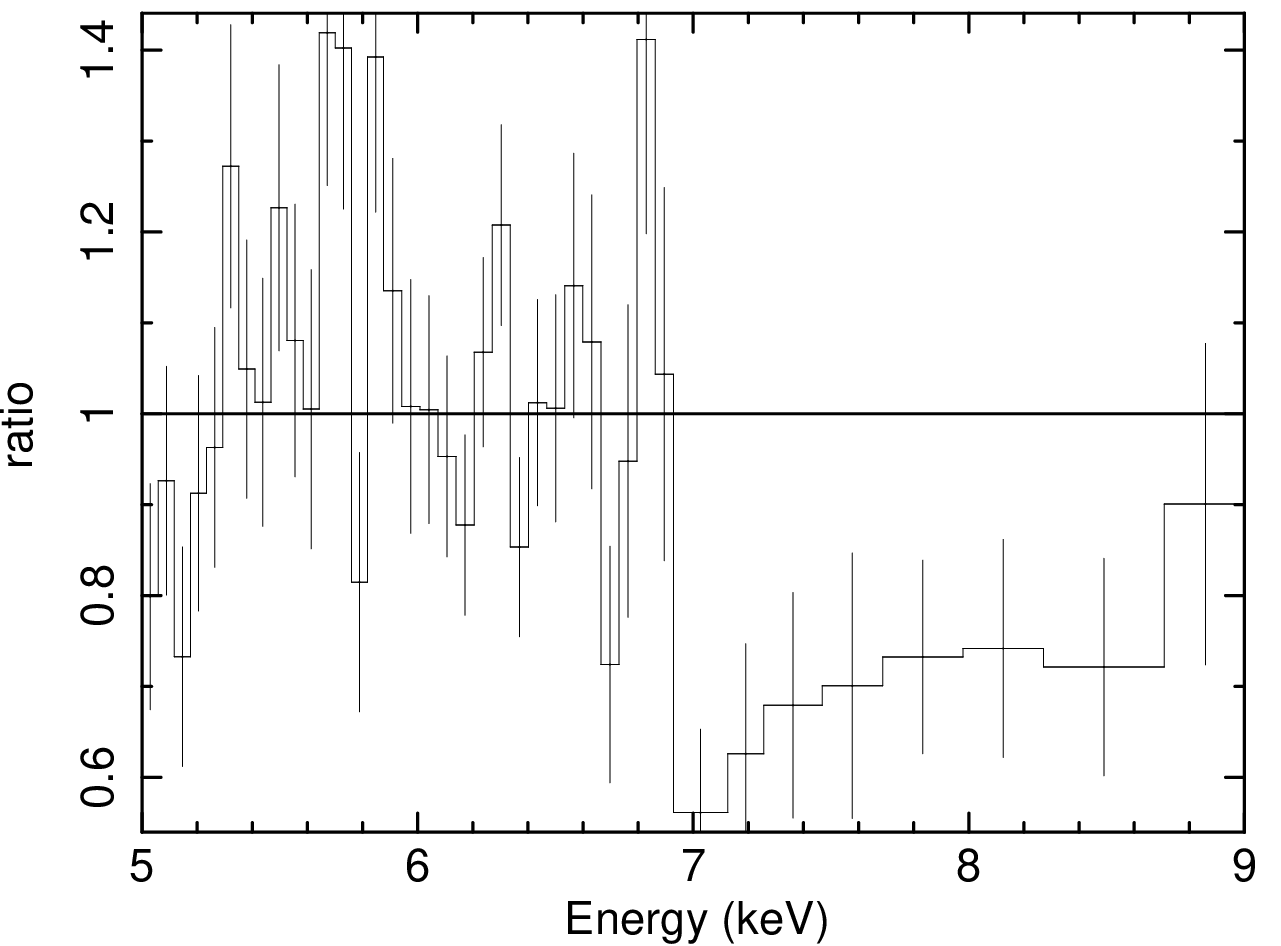}}}
\caption{Data/model ratio residuals from {\it Suzaku} observations of NGC 1194 (top left), NGC 1365 (2010 July 15, top right) and MCG-3-34-64 (below left). For each observation, the 5-7 keV data were fit with a powerlaw and a Gaussian component at 6.4 keV. The model was extrapolated to 9 keV, and the 5-9 keV data were divided by the folded model. Each spectrum shows a sharp edge at $\sim$ 7 keV.}
\label{fig:edge}
\end{figure}

NGC 1194 (Sey 1.9) , NGC 1365 (Sey 1.8; X-ray data from 2010 July 15) and MCG 3-34-64 are the three observational datasets in our sample showing the largest hardness ratios. These datasets show a sharp Fe K absorption edge at $\sim$ 7 keV (Figure~\ref{fig:edge}).  To determine whether the edges were significantly blurred, we fitted the 5-9 keV band of each with a powerlaw modified by the Galactic column. We parameterized the 6.4 keV Fe K$\alpha$ emission line with a Gaussian component and modeled the Fe K absorption edge near 7 keV with a smeared edge component (smedge in {\sc xspec}, \citealt{Ebisawa:1994qy}). All the parameters were allowed to vary, including the edge threshold energy, although this was not allowed to fall below the threshold energy of 7.11 keV for neutral Fe in the rest-frame. At 90\% confidence, the depth and width of the edge for NGC 1194, NGC 1365 (2010 July 15) and MCG 3-34-64 were $\tau$ = 1.3 $\pm$ 0.40 and $\sigma$ $<$ 0.15 keV, $\tau$ =  0.87 $\pm$ 0.10 and $\sigma$ $<$ 0.14 keV, and $\tau$ = 0.95 $\pm$ 0.30 and $\sigma$ $<$ 0.03 keV, respectively, showing that the absorption features are narrow and not likely to be reprocessed light from a blurred reflector.  

We performed an additional test for NGC 1365 as it had the highest quality data. We fitted the 6-9 keV bandpass using a powerlaw continuum with a Gaussian component for the Fe K$\alpha$ emission, having a fixed intrinsic width $\sigma =10$eV. We then allowed blurring into the model, using the {\sc xspec} model {\sc kdblur}, with emissivity index $q=3$, inclination angle $\theta=, 60^{\rm o}$ and the outer radius of the reflector at  $r_{out}=400 r_g$.  The data also required inclusion of absorption, which, in this limited bandpass, was adequately described by full covering from neutral gas with N$_H$ $\sim$ 10$^{23}$ cm$^{-2}$. With this model, we found the best fit  for the inner radius gave $r_{in}\sim$ 350 r$_g$. Stepping the inner radius down to  $r_{in}$=20 r$_g$ found no other minima as the inner radius was decreased and the fit at $r_{in}$=20 $r_g$ was statistically worse by  $\Delta\chi^2$ of 190 (F-test probability of $\sim$ 10$^{-27}$), compared to the best fit. This test confirms that the reprocessor  does not exhibit a blurred signature in this source. 

The most extreme sources possess sharp Fe K absorption edges. The absorption or reflection components that account for those edges also explain the hard spectral form of those sources. No relativistically-blurred components are required to explain these extreme sources, or any of the sample properties. We conclude that there is no evidence for any significant contribution by blurred  reflection to the  hard spectral form of local Seyfert galaxies.  
\section{Discussion}
The {\it Suzaku} observations of 1H 0419-577 \citep{Turner:2009ys}, PDS 456 \citep{Reeves:2009zt} and NGC 1365 \citep{Risaliti:2009lr} were first to reveal a marked 'hard excess' of flux above 10 keV. These results motivated an exploratory study of the hard excess phenomenon in the local type 1 AGN population. 

We have selected all type 1 AGN in the BAT 58-month catalog and cross-correlated them with the holdings of the {\it Suzaku} public archive. We have found the hard excess phenomenon to be a ubiquitous property of type 1 AGN. Comparisons between the hardness ratio distribution and simple models suggested both neutral, partial-covering models and disk reflection models to be viable. Joint consideration of the spectral hardness and equivalent width of Fe K$\alpha$ emission allowed us to rule out simple disk reflection models. 

The MyTorus  toroidal gas distribution accounts for the span of observed hardness ratios and line equivalent widths. However, evidence from detailed studies of nearby sources indicates that the reprocessor most likely is clumpy. We therefore performed simulations of an ensemble of Compton-thick reprocessing clouds. Predictions from the MCRT were in striking agreement with both the spectral hardness  and the line equivalent width distributions. 

In the context of a simple absorption model assuming partial-covering by neutral gas, $\sim$ 80\% of the sample has hardness ratios consistent with $>$ 50\% covering of the continuum by  Compton-thick material.  Furthermore, the model predictions for the MCRT are within the 1$\sigma$ errors of 85\% of the line equivalent widths in our sample. We conclude that both the spectral hardness and line equivalent width distribution strongly favor clumpy model interpretations. Our MCRT geometry shows excellent agreement with our observational results.  
 
\subsection{Gas location}
\subsubsection{Line width arguments}
Constraints obtained for the radial location of the absorbing gas are model dependent. Detailed analysis of the core of the Fe K$\alpha$ emission line and spectral variability may place the tightest constraints on the location of the circumnuclear gas in the context of the MCRT. \citet[][hereafter S10]{Shu:2010qy} studied the cores of the Fe K$\alpha$ lines of 36 type 1 AGN using {\it Chandra} HEG (high-energy grating) data. The spectral resolution of HEG is unparalleled in the Fe K band-- $\sim$ 39 eV at 6.4 keV, a factor of $\sim$ 4 better than the spectral resolution of {\it Suzaku} and {\it XMM-Newton}. S10 measured the FWHM of the Fe K$\alpha$ core, when the data quality was high enough, and found the mean FWHM of the Fe K$\alpha$ core to be 2060 $\pm$ 230 km s$^{-1}$. When comparing the FWHM of the core of Fe K emission  to the H$\beta$ optical emission line width, S10 found there to be no universal location of the Fe K$\alpha$ emission relative to the BLR, with possible locations ranging from $\sim$ 0.7-2 times the radius of the BLR out to the putative torus.  

We compared the FWHM Fe K core measurements and the corresponding H$\beta$ measurements of S10 for those sources common to both S10 and our study. On average, the FWHM measurements of the Fe K$\alpha$ core (6100$^{+30490}_{-1320}$ km s$^{-1}$) of the overlapping sample are comparable to the H$\beta$ FWHM ($\sim$ 4000 km s$^{-1}$), suggesting that the X-ray line-emitting clouds may be located within the optical broad line region. The broader component of Fe K$\alpha$ emission observed in some objects in our sample may arise from reflection off material from the inner parts of a Compton-thick accretion-disk wind \citep[e.g.,][]{Sim:2008ys, Tatum:2012fk}. The mean $\sigma$ width of the broader component was $\sigma$ $\sim$ 350eV, consistent with the findings of Patrick et al. (2012, MNRAS, submitted).

\subsubsection{Spectral variability arguments}
In the context of absorption models, some of the best studied sources have had their  spectral variability attributed to  variations in covering-fraction of the absorber \citep[i.e.,][]{Turner:2008qy, Miller:2008kx}. In such a picture, the location of the absorber must be consistent with the known time scales for spectral variability. 

Some constraints on gas location have been obtained for sources in our sample. For example, our  {\it Suzaku} observations of NGC 3227 show significant spectral variability on times scales $\sim$ 5 days (see the hardness ratio variations in Table~\ref{tab:table}), consistent with a virialized cloud location at the radius of the BLR. 

In a study of NGC 1365,  \citet{Risaliti:2009bh} compared variations in X-ray hardness,  Flux$_{2-5\, keV}$/ Flux$_{7-10\, keV}$, across a 60 ks {\it XMM} observation, in an attempt to isolate variations in the column of the X-ray absorber.  They split the data into three time intervals of $\sim 20$ ks, based on  hardness ratio variations.   Using these three time intervals in their time-resolved spectroscopy, they found that the source occupies two different flux states. The spectral  variations were attributed  to the passage of a cloud having column density $N_H \sim 3 \times$  10$^{23}$ cm$^{-2}$ crossing the line-of-sight. Assuming the cloud was in a Keplerian orbit and  cloud density was $n=10^{10}$ cm$^{-3}$, the timescale for the event yielded a radial location for the gas   $r \sim$ 10$^{16}$ cm \citep{Risaliti:2009bh}, consistent with the radial location and cloud density of the BLR. 

In general, there is irrefutable evidence from X-ray grating spectroscopy  that the absorber is outflowing \cite[e.g.][]{Turner:2008qy}. Consequently,  the assumption of Keplerian motion is surely inaccurate, leading to unreliable estimates of the gas radial location.   Independent evidence for the location of the X-ray reprocessor comes from recent work on X-ray reverberation. Estimates from sources such as NGC 4051 \citep{Miller:2010fj} and Ark 564 \citep{Legg:2012lr} suggest that reverberation occurs from material tens to hundreds of  gravitational radii from the central source.  However, the  reverberating gas may be highly ionized (e.g., \citealt{Legg:2012lr}) and may not be the same material causing the extreme spectral hardness noted here. 

\subsection{Implications for the intrinsic X-ray luminosity}

With the realization that large columns of gas cover a significant fraction of  our line-of-sight, we must consider the implications of this Compton-thick absorption to understand  the true nuclear continuum luminosity. For absorbers with such high column densities as found here, Compton scattering has a significant effect and must be taken into account. However, simple estimations of the Compton-scattering losses and absorption suppression of the continuum can overestimate the effect. The correction factor depends critically on the geometry of the gas, as we need to account for light scattered back into the line-of-sight by the cloud complex. 

We can estimate the correction factor between the observed and intrinsic X-ray luminosity in the context of the MCRT. For example, the observed 0.5--50 keV luminosity of MCG-03-34-64 (the third hardest source in our sample) is L$_x$ $\sim$ 1.7 x 10$^{43}$ erg s$^{-1}$. After correcting for scattering and absorption in the MCRT, the implied intrinsic luminosity is L$_{int}$(0.5 -- 50 keV) $\sim$ 1.7 x 10$^{44}$ erg s$^{-1}$, an order of magnitude higher than that observed. Importantly however, the calculated intrinsic luminosity does not exceed the Eddington luminosity, L$_{Edd}$ $\sim$ 10$^{45}$ erg s$^{-1}$ \citep{Miniutti:2007zr}, suggesting that, in general, partial-covering by a distribution of Compton-thick clouds is consistent with the energy budget of the system.

The correction required to understand the X-ray luminosity does not generally have a significant effect on the bolometric luminosity as the X-ray photons that are hidden from direct view are reprocessed into another bandpass and ultimately measured there.

\subsection{Comparison with other hard X-ray selected samples} 

Our work is not the first to determine global properties of AGN using a hard X-ray selected sample.  \citet{Winter:2009lq} (hereafter W09) examined the X-ray properties of local AGN in the BAT 9-month catalog  to determine the fraction of ``hidden'' AGN, which they defined, in part, as sources with scattered or leaking light fractions $<$ 3\%. Twenty-four percent of AGN in their sample were ``hidden'',  most likely by a geometrically-thick torus, and that the type 1 AGN population in their sample were relatively unobscured, with low column density gas (log N$_H$ $< $ 21.8) along the line-of-sight. In our sample, the type 1 AGN show significantly higher fractions of (unabsorbed) leaking light (2\%--97\%) and hence do not meet the W09 criterion for being hidden. Despite not meeting the W09 criterion for being hidden, the majority of sources selected in our sample do contain a significant amount of Compton-thick, clumpy absorption, a conclusion significantly different to that of W09.

\begin{figure}
\epsscale{.5}
\plotone{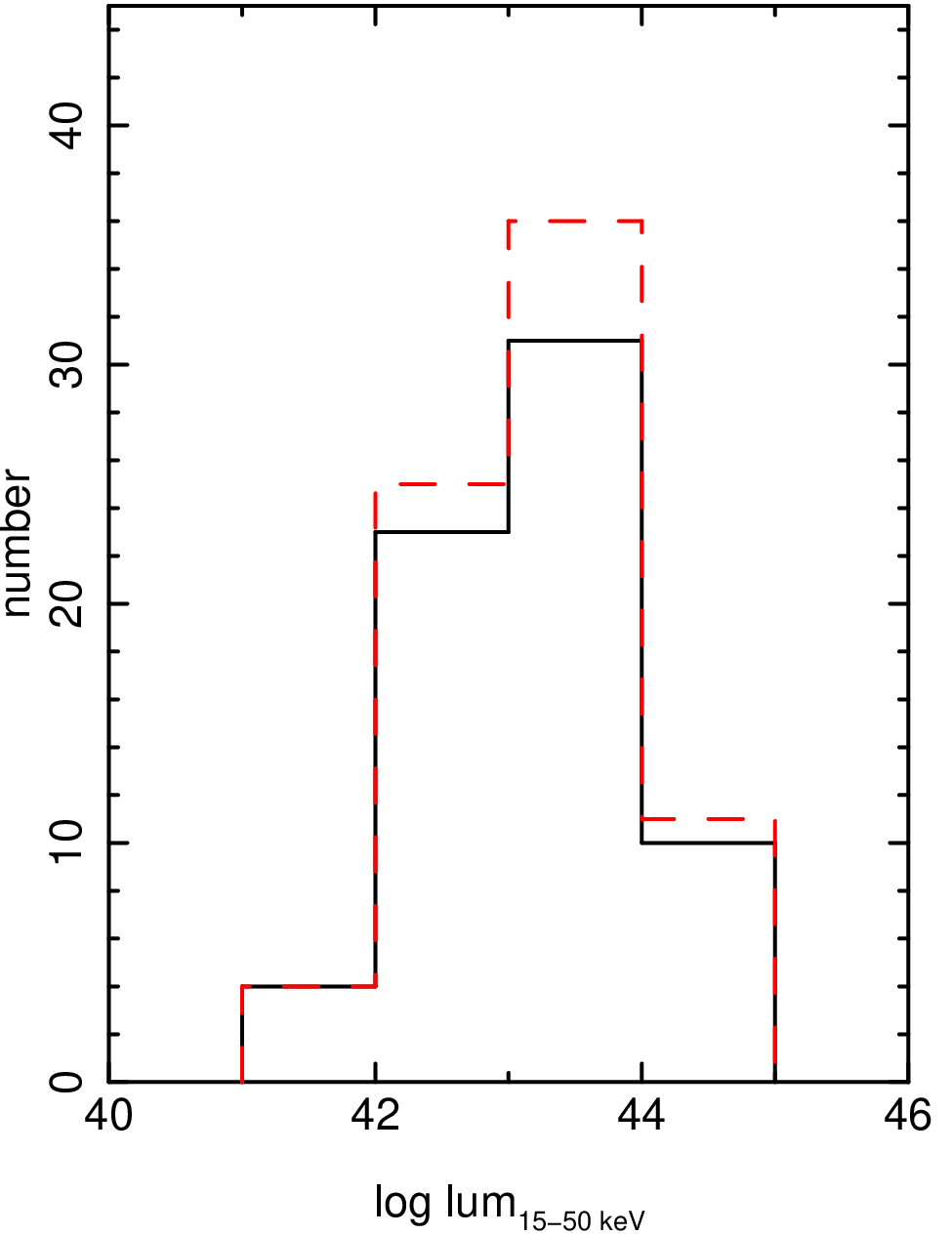}
\caption{The 15--50 keV luminosity distribution for those sources in our sample that are contained in the 9-month (solid black) and 58-month (dash red) BAT catalog.}
\label{fig:9_vs_58}
\end{figure}

Owing to the differences in our conclusions, we briefly review the most important differences in the samples and methodology. The W09 sample included all sources, both type 1 and type 2 AGN, from the 9-month BAT catalog \citep{Tueller:2008qq}. In contrast, our sample was derived from the 58-month BAT catalog, with AGN detection having substantially increased, by about a factor of 4, since the 9-month catalog. Further to this, we cross-correlated the sources in the 58-month catalog with those in the {\it Suzaku} archive and excluded all type 2 AGN, radio-loud AGN, and known LINERS. We also note that our AGN type criteria results in $\sim$ 20\% more sources (35 in W09, 43 in this work) when applied to the 58-month BAT catalog, than if we had applied it to the 9-month BAT catalog. Even though the 58-month BAT catalog probes deeper flux levels than the 9-month catalog, our sample is contained in the same observed luminosity parameter space (41.5 $<$ log lum$_{15-50 keV}$ $<$ 45) as the 9-month study (Figure~\ref{fig:9_vs_58}), suggesting that the flux bias in our sample selection is negligible.

There are important methodological differences between the study presented here, and that of W09. The spectral analysis of W09 was not restricted to the use of simultaneous data from the medium (2 -- 10 keV) and hard ($>$ 10 keV) X-ray bands, instead W09 used non-simultaneous data from {\it Swift XRT, ASCA SIS0, XMM-Newton pn, Chandra ACIS-S, and {\it Suzaku} XIS} in their spectral fitting. Most notably, the spectral fitting performed by W09 was restricted to data below 10 keV. Only utilizing the bandpass below 10 keV to detect broadband absorption would not be sensitive to Compton-thick absorption along the line-of-sight and may result in an underestimation of the intrinsic luminosity of the source. W09 compared the absorption-corrected flux of each source to the corresponding BAT flux (14 -- 195 keV) and found the relationship to be linear, taking this as validation of the absorption correction applied to the observed luminosity. However, comparing the data below 10 keV to the BAT flux assumes that the time-averaged BAT flux is representative of the flux state above 10 keV. Such an assumption is not consistent with the known variability above 10 keV seen in some AGN \citep[e.g.][]{Miniutti:2007jl, Miller:2008kx}. 

Broadband spectral analysis in the X-ray band offers better constraints on the intrinsic nature of AGN, especially for those objects with high column density clouds along the line-of-sight, as the model must be self-consistent with data below and above 10 keV. In our analysis, the role of {\it Suzaku} has been crucial, because it is essential to demonstrate the hard excess with an instrument that simultaneously observes the hard and soft X-ray bands:  the restriction to use only  {\it Suzaku} data for a BAT-selected sample is a key feature of our study. 

\citet[][hereafter W12]{Winter:2012qe} also selected a sample of sources from the 9-month BAT catalog, in order to understand the properties of the AGN warm absorbers. Their sample comprised 48 type 1 AGN, including both radio-loud and radio-quiet sources up to type 1.5. In this case, they used the simultaneous data in the {\it Suzaku} archive, where available. For those sources not observed by {\it Suzaku}, {\it XMM-Newton} spectra and time-averaged {\it Swift} BAT spectra were used for the broadband fits. Their model construction consisted of a powerlaw representing the continuum source, a blackbody model parameterizing emission below 2 keV, a Gaussian component modeling the Fe K$\alpha$ emission, and a distant reflector. The type 1 AGN were unobscured, consistent with W09, and that the line equivalent width of the Fe K$\alpha$ emission averaged $\sim$ 100 eV (consistent with W09 and our result). The relative strength of the distant reflector spanned 0 -- 5 times that expected from a standard thin disk of neutral material subtending 2$\pi$ sr at the continuum source. Comparing the overlapping objects of W12 and our study (excluding all sources with unconstrained relative reflection strength in W12), the average relative strength of the reflector in W12 is 2.10$^{+0.41}_{-0.44}$ times that expected from a standard thin disk of neutral material subtending 2$\pi$ sr at the continuum source. This result is similar to our observational finding in that the sample is harder than expected, especially for a sample of type 1 AGN. The wider span of the relative reflection strength seen in the ensemble result of W12 is not only, presumably, a consequence of including both radio-loud and radio-quiet AGN in the sample, but also, in part, a consequence of using non-simultaneous data. Inclusion of the of radio-loud in a comparison study is problematic, as the nuclear properties and the radio jet cannot be separated in the X-ray data. Our study is robust in that we identify general properties of a single subset of AGN over a broad energy range using simultaneous data.                 

\subsection{Comparison to the cosmic X-ray background} 
The 1 Ms {\it Chandra} Deep Field South \citep{Giacconi:2002ve}, 2 Ms {\it Chandra} Deep Field North \citep{Alexander:2003ul} and 800 ks {\it XMM-Newton} Lockman Hole \citep{Hasinger:2004qf} have confirmed the prediction of population synthesis models that the cosmic X-ray background originates from the integrated emission of extra-galactic point-like sources \citep{Gilli:2004kx}. Obscured AGN contribute to the cosmic X-ray background emission with most of the energy density being produced above 10 keV \citep{Comastri:2005yq,Gilli:2007qy,Treister:2009qy}. Our observational result suggests that a significant fraction of the type 1 AGN population is obscured with at least 50\% partial covering by Compton-thick material; therefore, it is interesting to compare a model representative of our mean spectrum to the spectral shape of the observed cosmic X-ray background. 

\begin{figure}
\epsscale{.5}
\plotone{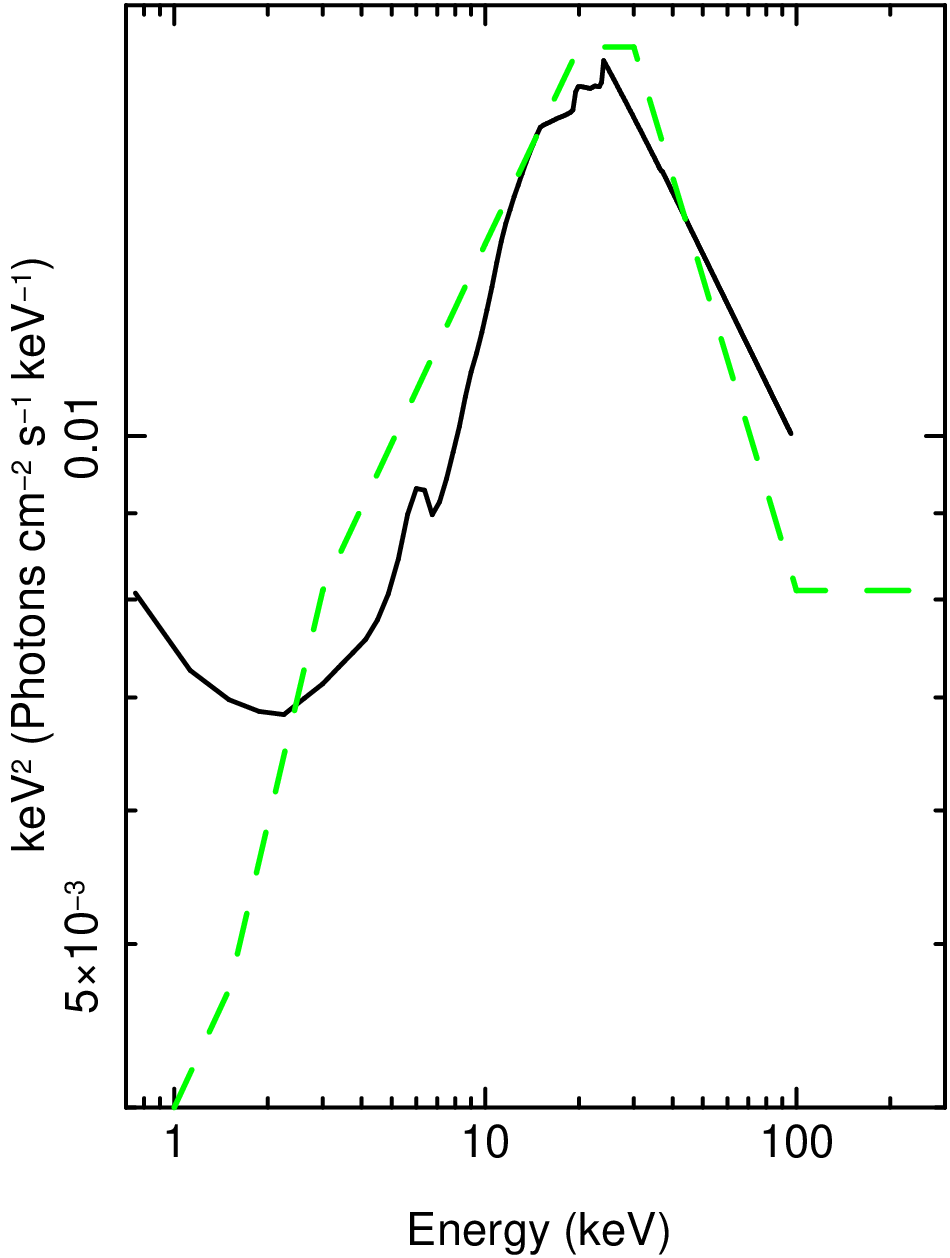}
\caption{ The best-fitting model for 1H 0419-577 (solid line), based on the model parameters from \citet{Turner:2009ys}, compared with the renormalized cosmic X-ray background data (dashed line) based on Figure 1 of \citet{Comastri:2005yq}.}
\label{fig:cxrb}
\end{figure}

As stated earlier, the hardness ratio measured for 1H 0419-577 was consistent with the weighted mean of the sample and may represent typical absorber parameters. We assumed 1H 0419-577 was representative of the sample and used the model construction and parameters from \citet{Turner:2009ys}. We compared the renormalized cosmic X-ray background data based on Figure 1 of \citet{Comastri:2005yq} to the spectral shape of 1H 0419-577 (see Figure~\ref{fig:cxrb}). We found the spectral shape of 1H 0419-577 to be consistent with the X-ray background above 2 keV with both having a characteristic peak $\sim$ 20 -- 30 keV.  Below 2 keV, the curve for 1H 0419-577 shows a marked rise in flux as compared to the X-ray background. These differences are to be expected considering that emission below 2 keV in type 1 AGN varies with covering fraction \citep{Miller:2008kx} and contributions to the X-ray background must include emission from both type 1 and type 2 AGN. 

X-ray population synthesis models typically have source redshifts of about unity. While it appears that adding contributions from objects like 1H 0419-577 and shifting them to $z=1$, would improve the agreement with the X-ray background, any more sophisticated analysis would require a proper consideration of the entire local AGN population and its cosmological evolution, which is beyond the scope of this paper. 

\section{Conclusion}
We conducted an exploratory study of the hard excess phenomenon in the local type 1 AGN population using sources from the Swift BAT 58-month catalog cross-correlated with the holdings of the {\it Suzaku} public archive. We extracted the hardness ratio and the equivalent width of the total Fe K$\alpha$ emission for each observation and found  that a simple disk reflection model does not provide an adequate representation of the sample results. Further, the presence of 
 sharp edges in the three hardest sources demonstrate that the extreme hardness cannot be dominated by blurred reflection components. 
 
However, a model comprising partial-covering by an ensemble of Compton-thick clouds is consistent with the observational result.  In the context of such a 
model, $\sim$ 80\% of the sample had a hardness ratio consistent with $>$ 50\% covering of the continuum by low-ionization, Compton-thick gas. Simple considerations suggested that the absorber lies at or within the optical BLR. Future work will include the results of our detailed analysis of this sample and a comparison of the results of this work with model predictions from a ionized, Compton-thick, partial-covering absorber, that includes Compton-scattering. More extensive spectral analysis using a parameterization of the Monte-Carlo radiative transfer model can further restrict the gas location as well as offer a better understanding of the geometry of the ensemble of Compton-thick clouds.       

\section{Acknowledgements}
MMT thanks Adam Patrick and Jason Gofford for aiding in the data reduction. MMT would like to acknowledge NASA grant NNX08AJ41G. TJT would like to acknowledge NASA grant  NNX11AJ57G.  LM acknowledges support from STFC grant  ST/H002456/1. 

\bibliographystyle{apj}      
\bibliography{list} 

\begin{thebibliography}{68}
\expandafter\ifx\csname natexlab\endcsname\relax\def\natexlab#1{#1}\fi

\bibitem[{{Alexander} {et~al.}(2003){Alexander}, {Bauer}, {Brandt},
  {Schneider}, {Hornschemeier}, {Vignali}, {Barger}, {Broos}, {Cowie},
  {Garmire}, {Townsley}, {Bautz}, {Chartas}, \& {Sargent}}]{Alexander:2003ul}
{Alexander}, D.~M., {Bauer}, F.~E., {Brandt}, W.~N., {Schneider}, D.~P.,
  {Hornschemeier}, A.~E., {Vignali}, C., {Barger}, A.~J., {Broos}, P.~S.,
  {Cowie}, L.~L., {Garmire}, G.~P., {Townsley}, L.~K., {Bautz}, M.~W.,
  {Chartas}, G., \& {Sargent}, W.~L.~W. 2003, \aj, 126, 539

\bibitem[{{Anders} \& {Grevesse}(1989)}]{Anders:1989kx}
{Anders}, E. \& {Grevesse}, N. 1989, \gca, 53, 197

\bibitem[{{Baumgartner} {et~al.}(2010){Baumgartner}, {Tueller}, {Markwardt}, \&
  {Skinner}}]{Baumgartner:2010fk}
{Baumgartner}, W.~H., {Tueller}, J., {Markwardt}, C., \& {Skinner}, G. 2010, in
  Bulletin of the American Astronomical Society, Vol.~42, AAS/High Energy
  Astrophysics Division \#11, 675

\bibitem[{{Blustin} {et~al.}(2007){Blustin}, {Kriss}, {Holczer}, {Behar},
  {Kaastra}, {Page}, {Kaspi}, {Branduardi-Raymont}, \&
  {Steenbrugge}}]{Blustin:2007nx}
{Blustin}, A.~J., {Kriss}, G.~A., {Holczer}, T., {Behar}, E., {Kaastra}, J.~S.,
  {Page}, M.~J., {Kaspi}, S., {Branduardi-Raymont}, G., \& {Steenbrugge}, K.~C.
  2007, \aap, 466, 107

\bibitem[{{Blustin} {et~al.}(2005){Blustin}, {Page}, {Fuerst},
  {Branduardi-Raymont}, \& {Ashton}}]{Blustin:2005mz}
{Blustin}, A.~J., {Page}, M.~J., {Fuerst}, S.~V., {Branduardi-Raymont}, G., \&
  {Ashton}, C.~E. 2005, \aap, 431, 111

\bibitem[{{Boldt}(1987)}]{Boldt:1987kx}
{Boldt}, E. 1987, in IAU Symposium, Vol. 124, Observational Cosmology, ed.
  A.~{Hewitt}, G.~{Burbidge}, \& L.~Z. {Fang}, 611--615

\bibitem[{{Brightman} \& {Nandra}(2012)}]{Brightman:2012fr}
{Brightman}, M. \& {Nandra}, K. 2012, arXiv1202.1291

\bibitem[{{Comastri} {et~al.}(2005){Comastri}, {Gilli}, \&
  {Hasinger}}]{Comastri:2005yq}
{Comastri}, A., {Gilli}, R., \& {Hasinger}, G. 2005, Experimental Astronomy,
  20, 41

\bibitem[{{Comastri} {et~al.}(1995){Comastri}, {Setti}, {Zamorani}, \&
  {Hasinger}}]{Comastri:1995pd}
{Comastri}, A., {Setti}, G., {Zamorani}, G., \& {Hasinger}, G. 1995, \aap, 296,
  1

\bibitem[{{Crenshaw} \& {Kraemer}(1999)}]{Crenshaw:1999th}
{Crenshaw}, D.~M. \& {Kraemer}, S.~B. 1999, \apj, 521, 572

\bibitem[{{Crenshaw} {et~al.}(2003){Crenshaw}, {Kraemer}, \&
  {George}}]{Crenshaw:2003vn}
{Crenshaw}, D.~M., {Kraemer}, S.~B., \& {George}, I.~M. 2003, \araa, 41, 117

\bibitem[{{Dickey} \& {Lockman}(1990)}]{Dickey:1990uq}
{Dickey}, J.~M. \& {Lockman}, F.~J. 1990, \araa, 28, 215

\bibitem[{{Ebisawa} {et~al.}(1994){Ebisawa}, {Ogawa}, {Aoki}, {Dotani},
  {Takizawa}, {Tanaka}, {Yoshida}, {Miyamoto}, {Iga}, {Hayashida}, {Kitamoto},
  \& {Terada}}]{Ebisawa:1994qy}
{Ebisawa}, K., {Ogawa}, M., {Aoki}, T., {Dotani}, T., {Takizawa}, M., {Tanaka},
  Y., {Yoshida}, K., {Miyamoto}, S., {Iga}, S., {Hayashida}, K., {Kitamoto},
  S., \& {Terada}, K. 1994, \pasj, 46, 375

\bibitem[{{George} \& {Fabian}(1991)}]{George:1991lr}
{George}, I.~M. \& {Fabian}, A.~C. 1991, \mnras, 249, 352

\bibitem[{{Giacconi} {et~al.}(2002){Giacconi}, {Zirm}, {Wang}, {Rosati},
  {Nonino}, {Tozzi}, {Gilli}, {Mainieri}, {Hasinger}, {Kewley}, {Bergeron},
  {Borgani}, {Gilmozzi}, {Grogin}, {Koekemoer}, {Schreier}, {Zheng}, \&
  {Norman}}]{Giacconi:2002ve}
{Giacconi}, R., {Zirm}, A., {Wang}, J., {Rosati}, P., {Nonino}, M., {Tozzi},
  P., {Gilli}, R., {Mainieri}, V., {Hasinger}, G., {Kewley}, L., {Bergeron},
  J., {Borgani}, S., {Gilmozzi}, R., {Grogin}, N., {Koekemoer}, A., {Schreier},
  E., {Zheng}, W., \& {Norman}, C. 2002, \apjs, 139, 369

\bibitem[{{Gilli}(2004)}]{Gilli:2004kx}
{Gilli}, R. 2004, Advances in Space Research, 34, 2470

\bibitem[{{Gilli} {et~al.}(2007){Gilli}, {Comastri}, \&
  {Hasinger}}]{Gilli:2007qy}
{Gilli}, R., {Comastri}, A., \& {Hasinger}, G. 2007, \aap, 463, 79

\bibitem[{{Gilli} {et~al.}(2010){Gilli}, {Vignali}, {Mignoli}, {Iwasawa},
  {Comastri}, \& {Zamorani}}]{Gilli:2010rt}
{Gilli}, R., {Vignali}, C., {Mignoli}, M., {Iwasawa}, K., {Comastri}, A., \&
  {Zamorani}, G. 2010, \aap, 519, A92

\bibitem[{{Haardt} \& {Maraschi}(1993)}]{Haardt:1993lr}
{Haardt}, F. \& {Maraschi}, L. 1993, \apj, 413, 507

\bibitem[{{Hasinger}(2004)}]{Hasinger:2004qf}
{Hasinger}, G. 2004, Nuclear Physics B Proceedings Supplements, 132, 86

\bibitem[{{Heckman} {et~al.}(2005){Heckman}, {Ptak}, {Hornschemeier}, \&
  {Kauffmann}}]{Heckman:2005rt}
{Heckman}, T.~M., {Ptak}, A., {Hornschemeier}, A., \& {Kauffmann}, G. 2005,
  \apj, 634, 161

\bibitem[{{Kallman} \& {Bautista}(2001)}]{Kallman:2001qf}
{Kallman}, T. \& {Bautista}, M. 2001, \apjs, 133, 221

\bibitem[{{Kaspi} {et~al.}(2002){Kaspi}, {Brandt}, {George}, {Netzer},
  {Crenshaw}, {Gabel}, {Hamann}, {Kaiser}, {Koratkar}, {Kraemer}, {Kriss},
  {Mathur}, {Mushotzky}, {Nandra}, {Peterson}, {Shields}, {Turner}, \&
  {Zheng}}]{Kaspi:2002lr}
{Kaspi}, S., {Brandt}, W.~N., {George}, I.~M., {Netzer}, H., {Crenshaw}, D.~M.,
  {Gabel}, J.~R., {Hamann}, F.~W., {Kaiser}, M.~E., {Koratkar}, A., {Kraemer},
  S.~B., {Kriss}, G.~A., {Mathur}, S., {Mushotzky}, R.~F., {Nandra}, K.,
  {Peterson}, B.~M., {Shields}, J.~C., {Turner}, T.~J., \& {Zheng}, W. 2002,
  \apj, 574, 643

\bibitem[{{Krolik} \& {Kallman}(1987)}]{Krolik:1987yq}
{Krolik}, J.~H. \& {Kallman}, T.~R. 1987, \apjl, 320, L5

\bibitem[{{LaMassa} {et~al.}(2011){LaMassa}, {Heckman}, {Ptak}, {Martins},
  {Wild}, {Sonnentrucker}, \& {Hornschemeier}}]{LaMassa:2011fj}
{LaMassa}, S.~M., {Heckman}, T.~M., {Ptak}, A., {Martins}, L., {Wild}, V.,
  {Sonnentrucker}, P., \& {Hornschemeier}, A. 2011, \apj, 729, 52

\bibitem[{{Lee} {et~al.}(2001){Lee}, {Ogle}, {Canizares}, {Marshall}, {Schulz},
  {Morales}, {Fabian}, \& {Iwasawa}}]{Lee:2001fk}
{Lee}, J.~C., {Ogle}, P.~M., {Canizares}, C.~R., {Marshall}, H.~L., {Schulz},
  N.~S., {Morales}, R., {Fabian}, A.~C., \& {Iwasawa}, K. 2001, \apjl, 554, L13

\bibitem[{{Legg} {et~al.}(2012){Legg}, {Miller}, {Turner}, {Giustini},
  {Reeves}, \& {Kraemer}}]{Legg:2012lr}
{Legg}, E., {Miller}, L., {Turner}, T.~J., {Giustini}, M., {Reeves}, J.~N., \&
  {Kraemer}, S.~B. 2012, arXiv:1210.0469

\bibitem[{{Maeda} {et~al.}(2008){Maeda}, {Someya}, {Ishida}, {the XRT team},
  {Hayashida}, {Mori}, \& the XIS~team}]{Maeda:2008ab}
{Maeda}, Y., {Someya}, K., {Ishida}, M., {the XRT team}, {Hayashida}, K.,
  {Mori}, H., \& the XIS~team. 2008, JX-ISAS-SUZAKU-MEMO-2008-06

\bibitem[{{Magdziarz} \& {Zdziarski}(1995)}]{Magdziarz:1995uq}
{Magdziarz}, P. \& {Zdziarski}, A.~A. 1995, \mnras, 273, 837

\bibitem[{{McKernan} \& {Yaqoob}(1998)}]{McKernan:1998vn}
{McKernan}, B. \& {Yaqoob}, T. 1998, \apjl, 501, L29

\bibitem[{{McKernan} {et~al.}(2007){McKernan}, {Yaqoob}, \&
  {Reynolds}}]{McKernan:2007oq}
{McKernan}, B., {Yaqoob}, T., \& {Reynolds}, C.~S. 2007, \mnras, 379, 1359

\bibitem[{{Miller} {et~al.}(2008){Miller}, {Turner}, \&
  {Reeves}}]{Miller:2008kx}
{Miller}, L., {Turner}, T.~J., \& {Reeves}, J.~N. 2008, \aap, 483, 437

\bibitem[{{Miller} {et~al.}(2009){Miller}, {Turner}, \&
  {Reeves}}]{Miller:2009cr}
---. 2009, \mnras, 399, L69

\bibitem[{{Miller} {et~al.}(2007){Miller}, {Turner}, {Reeves}, {George},
  {Kraemer}, \& {Wingert}}]{Miller:2007fe}
{Miller}, L., {Turner}, T.~J., {Reeves}, J.~N., {George}, I.~M., {Kraemer},
  S.~B., \& {Wingert}, B. 2007, \aap, 463, 131

\bibitem[{{Miller} {et~al.}(2010){Miller}, {Turner}, {Reeves}, {Lobban},
  {Kraemer}, \& {Crenshaw}}]{Miller:2010fj}
{Miller}, L., {Turner}, T.~J., {Reeves}, J.~N., {Lobban}, A., {Kraemer}, S.~B.,
  \& {Crenshaw}, D.~M. 2010, \mnras, 403, 196

\bibitem[{{Miniutti} {et~al.}(2007{\natexlab{a}}){Miniutti}, {Fabian},
  {Anabuki}, {Crummy}, {Fukazawa}, {Gallo}, {Haba}, {Hayashida}, {Holt},
  {Kunieda}, {Larsson}, {Markowitz}, {Matsumoto}, {Ohno}, {Reeves},
  {Takahashi}, {Tanaka}, {Terashima}, {Torii}, {Ueda}, {Ushio}, {Watanabe},
  {Yamauchi}, \& {Yaqoob}}]{Miniutti:2007jl}
{Miniutti}, G., {Fabian}, A.~C., {Anabuki}, N., {Crummy}, J., {Fukazawa}, Y.,
  {Gallo}, L., {Haba}, Y., {Hayashida}, K., {Holt}, S., {Kunieda}, H.,
  {Larsson}, J., {Markowitz}, A., {Matsumoto}, C., {Ohno}, M., {Reeves}, J.~N.,
  {Takahashi}, T., {Tanaka}, Y., {Terashima}, Y., {Torii}, K., {Ueda}, Y.,
  {Ushio}, M., {Watanabe}, S., {Yamauchi}, M., \& {Yaqoob}, T.
  2007{\natexlab{a}}, \pasj, 59, 315

\bibitem[{{Miniutti} {et~al.}(2007{\natexlab{b}}){Miniutti}, {Ponti}, {Dadina},
  {Cappi}, \& {Malaguti}}]{Miniutti:2007zr}
{Miniutti}, G., {Ponti}, G., {Dadina}, M., {Cappi}, M., \& {Malaguti}, G.
  2007{\natexlab{b}}, \mnras, 375, 227

\bibitem[{{Murphy} \& {Yaqoob}(2009)}]{Murphy:2009lq}
{Murphy}, K.~D. \& {Yaqoob}, T. 2009, \mnras, 397, 1549

\bibitem[{{Nandra} \& {George}(1994)}]{Nandra:1994fj}
{Nandra}, K. \& {George}, I.~M. 1994, \mnras, 267, 974

\bibitem[{{Nandra} {et~al.}(2007){Nandra}, {O'Neill}, {George}, \&
  {Reeves}}]{Nandra:2007lr}
{Nandra}, K., {O'Neill}, P.~M., {George}, I.~M., \& {Reeves}, J.~N. 2007,
  \mnras, 382, 194

\bibitem[{{Netzer} {et~al.}(2003){Netzer}, {Kaspi}, {Behar}, {Brandt},
  {Chelouche}, {George}, {Crenshaw}, {Gabel}, {Hamann}, {Kraemer}, {Kriss},
  {Nandra}, {Peterson}, {Shields}, \& {Turner}}]{Netzer:2003wd}
{Netzer}, H., {Kaspi}, S., {Behar}, E., {Brandt}, W.~N., {Chelouche}, D.,
  {George}, I.~M., {Crenshaw}, D.~M., {Gabel}, J.~R., {Hamann}, F.~W.,
  {Kraemer}, S.~B., {Kriss}, G.~A., {Nandra}, K., {Peterson}, B.~M., {Shields},
  J.~C., \& {Turner}, T.~J. 2003, \apj, 599, 933

\bibitem[{{Reeves} {et~al.}(2004){Reeves}, {Nandra}, {George}, {Pounds},
  {Turner}, \& {Yaqoob}}]{Reeves:2004qy}
{Reeves}, J.~N., {Nandra}, K., {George}, I.~M., {Pounds}, K.~A., {Turner},
  T.~J., \& {Yaqoob}, T. 2004, \apj, 602, 648

\bibitem[{{Reeves} {et~al.}(2009){Reeves}, {O'Brien}, {Braito}, {Behar},
  {Miller}, {Turner}, {Fabian}, {Kaspi}, {Mushotzky}, \&
  {Ward}}]{Reeves:2009zt}
{Reeves}, J.~N., {O'Brien}, P.~T., {Braito}, V., {Behar}, E., {Miller}, L.,
  {Turner}, T.~J., {Fabian}, A.~C., {Kaspi}, S., {Mushotzky}, R., \& {Ward}, M.
  2009, \apj, 701, 493

\bibitem[{{Risaliti} {et~al.}(2005){Risaliti}, {Bianchi}, {Matt}, {Baldi},
  {Elvis}, {Fabbiano}, \& {Zezas}}]{Risaliti:2005uq}
{Risaliti}, G., {Bianchi}, S., {Matt}, G., {Baldi}, A., {Elvis}, M.,
  {Fabbiano}, G., \& {Zezas}, A. 2005, \apjl, 630, L129

\bibitem[{{Risaliti} {et~al.}(2009{\natexlab{a}}){Risaliti}, {Braito},
  {Laparola}, {Bianchi}, {Elvis}, {Fabbiano}, {Maiolino}, {Matt}, {Reeves},
  {Salvati}, \& {Wang}}]{Risaliti:2009lr}
{Risaliti}, G., {Braito}, V., {Laparola}, V., {Bianchi}, S., {Elvis}, M.,
  {Fabbiano}, G., {Maiolino}, R., {Matt}, G., {Reeves}, J., {Salvati}, M., \&
  {Wang}, J. 2009{\natexlab{a}}, \apjl, 705, L1

\bibitem[{{Risaliti} {et~al.}(2009{\natexlab{b}}){Risaliti}, {Miniutti},
  {Elvis}, {Fabbiano}, {Salvati}, {Baldi}, {Braito}, {Bianchi}, {Matt},
  {Reeves}, {Soria}, \& {Zezas}}]{Risaliti:2009bh}
{Risaliti}, G., {Miniutti}, G., {Elvis}, M., {Fabbiano}, G., {Salvati}, M.,
  {Baldi}, A., {Braito}, V., {Bianchi}, S., {Matt}, G., {Reeves}, J., {Soria},
  R., \& {Zezas}, A. 2009{\natexlab{b}}, \apj, 696, 160

\bibitem[{{Scott} {et~al.}(2011){Scott}, {Stewart}, {Mateos}, {Alexander},
  {Hutton}, \& {Ward}}]{Scott:2011fk}
{Scott}, A.~E., {Stewart}, G.~C., {Mateos}, S., {Alexander}, D.~M., {Hutton},
  S., \& {Ward}, M.~J. 2011, \mnras, 417, 992

\bibitem[{{Shu} {et~al.}(2010){Shu}, {Yaqoob}, \& {Wang}}]{Shu:2010qy}
{Shu}, X.~W., {Yaqoob}, T., \& {Wang}, J.~X. 2010, \apjs, 187, 581

\bibitem[{{Sim} {et~al.}(2008){Sim}, {Long}, {Miller}, \&
  {Turner}}]{Sim:2008ys}
{Sim}, S.~A., {Long}, K.~S., {Miller}, L., \& {Turner}, T.~J. 2008, \mnras,
  388, 611

\bibitem[{{Sim} {et~al.}(2010){Sim}, {Miller}, {Long}, {Turner}, \&
  {Reeves}}]{Sim:2010fr}
{Sim}, S.~A., {Miller}, L., {Long}, K.~S., {Turner}, T.~J., \& {Reeves}, J.~N.
  2010, \mnras, 404, 1369

\bibitem[{{Tanaka} {et~al.}(2004){Tanaka}, {Boller}, {Gallo}, {Keil}, \&
  {Ueda}}]{Tanaka:2004fk}
{Tanaka}, Y., {Boller}, T., {Gallo}, L., {Keil}, R., \& {Ueda}, Y. 2004, \pasj,
  56, L9

\bibitem[{{Tatum} {et~al.}(2012{\natexlab{a}}){Tatum}, {Turner}, {Miller}, \&
  {Reeves}}]{Tatum:2012inprep}
{Tatum}, M.~M., {Turner}, T.~J., {Miller}, L., \& {Reeves}, J.~N.
  2012{\natexlab{a}}, in prep.

\bibitem[{{Tatum} {et~al.}(2012{\natexlab{b}}){Tatum}, {Turner}, {Sim},
  {Miller}, {Reeves}, {Patrick}, \& {Long}}]{Tatum:2012fk}
{Tatum}, M.~M., {Turner}, T.~J., {Sim}, S.~A., {Miller}, L., {Reeves}, J.~N.,
  {Patrick}, A.~R., \& {Long}, K.~S. 2012{\natexlab{b}}, \apj, 752, 94

\bibitem[{{Tombesi} {et~al.}(2011){Tombesi}, {Cappi}, {Reeves}, {Palumbo},
  {Braito}, \& {Dadina}}]{Tombesi:2011lr}
{Tombesi}, F., {Cappi}, M., {Reeves}, J.~N., {Palumbo}, G.~G.~C., {Braito}, V.,
  \& {Dadina}, M. 2011, ArXiv e-prints

\bibitem[{{Tombesi} {et~al.}(2010){Tombesi}, {Cappi}, {Reeves}, {Palumbo},
  {Yaqoob}, {Braito}, \& {Dadina}}]{Tombesi:2010yq}
{Tombesi}, F., {Cappi}, M., {Reeves}, J.~N., {Palumbo}, G.~G.~C., {Yaqoob}, T.,
  {Braito}, V., \& {Dadina}, M. 2010, \aap, 521, A57

\bibitem[{{Treister} {et~al.}(2009){Treister}, {Urry}, \&
  {Virani}}]{Treister:2009qy}
{Treister}, E., {Urry}, C.~M., \& {Virani}, S. 2009, \apj, 696, 110

\bibitem[{{Tueller} {et~al.}(2008){Tueller}, {Mushotzky}, {Barthelmy},
  {Cannizzo}, {Gehrels}, {Markwardt}, {Skinner}, \& {Winter}}]{Tueller:2008qq}
{Tueller}, J., {Mushotzky}, R.~F., {Barthelmy}, S., {Cannizzo}, J.~K.,
  {Gehrels}, N., {Markwardt}, C.~B., {Skinner}, G.~K., \& {Winter}, L.~M. 2008,
  \apj, 681, 113

\bibitem[{{Turner} \& {Miller}(2009)}]{Turner:2009lr}
{Turner}, T.~J. \& {Miller}, L. 2009, \aapr, 17, 47

\bibitem[{{Turner} {et~al.}(2011){Turner}, {Miller}, {Kraemer}, \&
  {Reeves}}]{Turner:2011uq}
{Turner}, T.~J., {Miller}, L., {Kraemer}, S.~B., \& {Reeves}, J.~N. 2011, \apj,
  733, 48

\bibitem[{{Turner} {et~al.}(2009){Turner}, {Miller}, {Kraemer}, {Reeves}, \&
  {Pounds}}]{Turner:2009ys}
{Turner}, T.~J., {Miller}, L., {Kraemer}, S.~B., {Reeves}, J.~N., \& {Pounds},
  K.~A. 2009, \apj, 698, 99

\bibitem[{{Turner} {et~al.}(2007){Turner}, {Miller}, {Reeves}, \&
  {Kraemer}}]{Turner:2007fj}
{Turner}, T.~J., {Miller}, L., {Reeves}, J.~N., \& {Kraemer}, S.~B. 2007, \aap,
  475, 121

\bibitem[{{Turner} {et~al.}(2008){Turner}, {Reeves}, {Kraemer}, \&
  {Miller}}]{Turner:2008qy}
{Turner}, T.~J., {Reeves}, J.~N., {Kraemer}, S.~B., \& {Miller}, L. 2008, \aap,
  483, 161

\bibitem[{{V{\'e}ron-Cetty} \& {V{\'e}ron}(2006)}]{Veron-Cetty:2006lr}
{V{\'e}ron-Cetty}, M.-P. \& {V{\'e}ron}, P. 2006, \aap, 455, 773

\bibitem[{{Walton} {et~al.}(2010){Walton}, {Reis}, \& {Fabian}}]{Walton:2010lr}
{Walton}, D.~J., {Reis}, R.~C., \& {Fabian}, A.~C. 2010, \mnras, 408, 601

\bibitem[{{Wilms} {et~al.}(2000){Wilms}, {Allen}, \& {McCray}}]{Wilms:2000qy}
{Wilms}, J., {Allen}, A., \& {McCray}, R. 2000, \apj, 542, 914

\bibitem[{{Winter} {et~al.}(2009){Winter}, {Mushotzky}, {Reynolds}, \&
  {Tueller}}]{Winter:2009lq}
{Winter}, L.~M., {Mushotzky}, R.~F., {Reynolds}, C.~S., \& {Tueller}, J. 2009,
  \apj, 690, 1322

\bibitem[{{Winter} {et~al.}(2012){Winter}, {Veilleux}, {McKernan}, \&
  {Kallman}}]{Winter:2012qe}
{Winter}, L.~M., {Veilleux}, S., {McKernan}, B., \& {Kallman}, T.~R. 2012,
  \apj, 745, 107

\bibitem[{{Yaqoob} {et~al.}(2010){Yaqoob}, {Murphy}, {Miller}, \&
  {Turner}}]{Yaqoob:2010fj}
{Yaqoob}, T., {Murphy}, K.~D., {Miller}, L., \& {Turner}, T.~J. 2010, \mnras,
  401, 411

\end{thebibliography}

\newpage
\begin{deluxetable}{lccccclcc}
\tabletypesize{\scriptsize}
\tablecaption{X-ray Observation Source List} 
\tablewidth{0pt}
\tablehead{
\colhead{Object} &
\colhead{ObsID}&
\colhead{RA (h m s)\tablenotemark{1}}&
\colhead{Dec ({$^\circ$ \arcmin\, \arcsec)}\tablenotemark{1}}&
\colhead{z\tablenotemark{1}}&
\colhead{N$_H$(Gal)\tablenotemark{2}}&
\colhead{Type\tablenotemark{1}}&
\colhead{Hardness Ratio\tablenotemark{3}}&
\colhead{Total EW (eV)\tablenotemark{4}}\\
}
\startdata
\multicolumn{9}{c}{}\\
1H 0419-577&702041010&04 26 00.7&--57 12 01&0.104&0.020&1.5&1.45$\pm{0.06}$&42$^{+1}_{-1}$\\
1H 0419-577&704064010&04 26 00.7&--57 12 01&0.104&0.020&1.5&1.53$\pm{0.07}$&20$^{+7}_{-9}$\\
2MASXJ09043699+5536025&704027010&09 04 36.9&+55 36 03&0.037&0.028&1&3.24$\pm{2.30}$&97$^{+5}_{-5}$\\
Ark 120&702014010\tablenotemark{5}&05 16 11.4&--00 08 59&0.033&0.126&1&1.10$\pm{0.05}$&197$^{+1}_{-1}$\\
ESO 362-18&703014010&05 19 35.8&--32 39 28&0.012&0.018&1.5&1.44$\pm{0.55}$&175$^{+1}_{-1}$\\
ESO 548-G081&704026010\tablenotemark{5}&03 42 03.7&--21 14 39&0.015&0.032&1\tablenotemark{6}&1.66$\pm{0.14}$&94$^{+1}_{-1}$\\
Fairall 9&702043010\tablenotemark{5}&01 23 45.8&--58 48 21&0.047&0.032&1.2&1.31$\pm{0.04}$&186$^{+1}_{-1}$\\
Fairall 9&705063010&01 23 45.8&--58 48 21&0.047&0.032&1.2&1.58$\pm{0.04}$&78$^{+9}_{-7}$\\
IC4329A&702113010\tablenotemark{5}&13 49 19.3&--30 18 34&0.016&0.044&1.2&1.57$\pm{0.04}$&115$^{+11}_{-15}$\\
IC4329A&702113030&13 49 19.3&--30 18 34&0.016&0.044&1.2&1.66$\pm{0.04}$&58$^{+10}_{-10}$\\
IC4329A&702113040\tablenotemark{5}&13 49 19.3&--30 18 34&0.016&0.044&1.2&1.49$\pm{0.04}$&126$^{+1}_{-1}$\\
IC4329A&702113050&13 49 19.3&--30 18 34&0.016&0.044&1.2&1.90$\pm{0.05}$&98$^{+14}_{-19}$\\
IC4329A&702113020&13 49 19.3&--30 18 34&0.016&0.044&1.2&1.43$\pm{0.03}$&50$^{+10}_{-11}$\\
IGR J16185-5928&702123010&16 18 36.4&--59 27 17&0.035&0.247&1 NL\tablenotemark{6}&1.67$\pm{0.15}$&75$^{+30}_{-22}$\\
IGR J19378-0617&705055010&19 37 33.0&--06 13 05&0.010&0.148&1.5&0.68$\pm{0.07}$&47$^{+15}_{-15}$\\
MCG+04-22-42&704028010&09 23 43.0&+22 54 33&0.032&0.034&1.2&1.27$\pm{0.14}$&95$^{+2}_{-2}$\\
MCG+08-11-11&702112010&05 54 53.6&+46 26 22&0.021&0.209&1.5&1.54$\pm{0.03}$&100$^{+9}_{-10}$\\
MCG-02-14-09&703060010\tablenotemark{5}&05 16 21.2&--10 33 41&0.029&0.093&1&1.44$\pm{0.25}$&200$^{+1}_{-2}$\\
MCG-02-58-22&704032010&23 04 43.5&--08 41 09&0.047&0.036&1.5&1.82$\pm{0.03}$&47$^{+8}_{-6}$\\
MCG-03-34-64&702112010&13 22 24.4&--16 34 42&0.017&0.058&1 h&7.53$\pm{0.56}$&372$^{+65}_{-75}$\\
MCG-06-30-15&100004010&13 35 53.7&--34 17 44&0.008&0.041&1.2&1.24$\pm{0.05}$&98$^{+1}_{-1}$\\
MCG-06-30-15&700007010&13 35 53.7&--34 17 44&0.008&0.041&1.2&1.33$\pm{0.03}$&80$^{+11}_{-11}$\\
MCG-06-30-15&700007020&13 35 53.7&--34 17 44&0.008&0.041&1.2&1.65$\pm{0.04}$&63$^{+11}_{-9}$\\
MCG-06-30-15&700007030&13 35 53.7&--34 17 44&0.008&0.041&1.2&1.59$\pm{0.03}$&49$^{+7}_{-6}$\\
MR 2251-178&704055010&22 54 05.8&--17 34 55&0.064&0.027&1.5&1.34$\pm{0.04}$&38$^{+7}_{-13}$\\
Mrk 1018&704044010&02 06 16.0&--00 17 29&0.042&0.026&1.9&1.73$\pm{0.14}$&76$^{+1}_{-2}$\\
Mrk 110&702124010&09 25 12.9&+52 17 11&0.035&0.015&1&1.36$\pm{0.06}$&64$^{+13}_{-18}$\\
Mrk 279&704031010&13 53 03.4&+69 18 30&0.031&0.018&1&3.21$\pm{0.21}$&234$^{+3}_{-3}$\\
Mrk 335&701031010\tablenotemark{5}&00 06 19.5&+20 12 10&0.026&0.040&1&1.05$\pm{0.09}$&181$^{+14}_{-12}$\\
Mrk 352&704025010&00 59 53.3&+31 49 37&0.015&0.055&1&1.42$\pm{0.28}$&89$^{+39}_{-37}$\\
Mrk 359&701082010&01 27 32.5&+19 10 44&0.017&0.048&1 NL&1.38$\pm{0.44}$&135$^{+29}_{-38}$\\
Mrk 509&701093010&20 44 09.7&-10 43 25&0.034&0.041&1.5&1.41$\pm{0.06}$&50$^{+17}_{-15}$\\
Mrk 509&701093020&20 44 09.7&-10 43 25&0.034&0.041&1.5&1.56$\pm{0.06}$&56$^{+18}_{-17}$\\
Mrk 509&701093030&20 44 09.7&-10 43 25&0.034&0.041&1.5&1.51$\pm{0.08}$&54$^{+19}_{-19}$\\
Mrk 509&701093040&20 44 09.7&-10 43 25&0.034&0.041&1.5&1.50$\pm{0.06}$&58$^{+16}_{-13}$\\
Mrk 766&701035010&12 18 26.5&+29 48 46&0.013&0.017&1 NL&1.27$\pm{0.11}$&63$^{+28}_{-33}$\\
Mrk 766&701035020&12 18 26.5&+29 48 46&0.013&0.017&1 NL&1.18$\pm{0.10}$&71$^{+25}_{-24}$\\
Mrk 79&702044010&07 42 32.8&+49 48 35&0.022&0.053&1.2&1.78$\pm{0.16}$&159$^{+19}_{-19}$\\
Mrk 841&701084010&15 04 01.2&+10 26 16&0.036&0.023&1.5&1.22$\pm{0.09}$&83$^{+1}_{-1}$\\
Mrk 841&701084020&15 04 01.2&+10 26 16&0.036&0.023&1.5&2.03$\pm{0.12}$&58$^{+20}_{-15}$\\
NGC 1194&704043010&03 03 49.1&--01 06 13&0.014&0.071&1.9&31.17$\pm{2.92}$&789$^{+121}_{-140}$\\
NGC 1365&702047010&03 33 36.4&--36 08 25&0.005&0.014&1.8&3.07$\pm{0.08}$&331$^{+15}_{-15}$\\
NGC 1365&705031010&03 33 36.4&--36 08 25&0.005&0.014&1.8&11.12$\pm{0.33}$&459$^{+27}_{-20}$\\
NGC 1365&705031020&03 33 36.4&--36 08 25&0.005&0.014&1.8&12.72$\pm{0.22}$&523$^{+29}_{-20}$\\
NGC 3227&703022010&10 23 30.6&+19 51 54&0.004&0.020&1.5&1.95$\pm{0.43}$&113$^{+14}_{-17}$\\
NGC 3227&703022020&10 23 30.6&+19 51 54&0.004&0.020&1.5&3.97$\pm{0.12}$&285$^{+20}_{-20}$\\
NGC 3227&703022030&10 23 30.6&+19 51 54&0.004&0.020&1.5&2.67$\pm{0.08}$&163$^{+16}_{-19}$\\
NGC 3227&703022040&10 23 30.6&+19 51 54&0.004&0.020&1.5&5.28$\pm{0.22}$&497$^{+33}_{-34}$\\
NGC 3227&703022050&10 23 30.6&+19 51 54&0.004&0.020&1.5&2.92$\pm{0.10}$&162$^{+13}_{-13}$\\
NGC 3227&703022060&10 23 30.6&+19 51 54&0.004&0.020&1.5&2.67$\pm{0.11}$&163$^{+24}_{-17}$\\
NGC 3516&100031010&11 06 47.5&+72 34 07&0.009&0.031&1.5&2.97$\pm{0.06}$&215$^{+14}_{-12}$\\
NGC 3516&704062010&11 06 47.5&+72 34 07&0.009&0.031&1.5&2.58$\pm{0.07}$&281$^{+25}_{-19}$\\
NGC 3783&701033010&11 39 01.7&--37 44 19&0.010&0.085&1.5&2.02$\pm{0.05}$&164$^{+10}_{-10}$\\
NGC 3783&704063010&11 39 01.7&--37 44 19&0.010&0.085&1.5&1.96$\pm{0.03}$&102$^{+6}_{-5}$\\
NGC 4051&700004010&12 03 09.6&+44 31 53&0.002&0.013&1&2.75$\pm{0.19}$&181$^{+18}_{-13}$\\
NGC 4051&703023010&12 03 09.6&+44 31 53&0.002&0.013&1&1.41$\pm{0.04}$&82$^{+9}_{-10}$\\
NGC 4051&703023020&12 03 09.6&+44 31 53&0.002&0.013&1&1.69$\pm{0.10}$&111$^{+18}_{-24}$\\
NGC 4151&701034010&12 10 32.6&+39 24 21&0.003&0.020&1.5&3.75$\pm{0.06}$&388$^{+19}_{-11}$\\
NGC 4593&702040010&12 39 39.4&--05 20 39&0.009&0.023&1&1.98$\pm{0.14}$&278$^{+19}_{-21}$\\
NGC 5506&701030010\tablenotemark{5}&14 13 14.9&--03 12 27&0.006&0.038&1 i&1.69$\pm{0.03}$&118$^{+7}_{-9}$\\
NGC 5506&701030020\tablenotemark{5}&14 13 14.9&--03 12 27&0.006&0.038&1 i&1.70$\pm{0.03}$&120$^{+7}_{-7}$\\
NGC 5506&701030030&14 13 14.9&--03 12 27&0.006&0.038&1 i&1.69$\pm{0.03}$&75$^{+9}_{-10}$\\
NGC 5548&702042010&14 17 59.5&+25 08 12&0.017&0.017&1.5&3.52$\pm{0.34}$&193$^{+6}_{-4}$\\
NGC 5548&702042020&14 17 59.5&+25 08 12&0.017&0.017&1.5&2.29$\pm{0.16}$&151$^{+3}_{-2}$\\
NGC 5548&702042040&14 17 59.5&+25 08 12&0.017&0.017&1.5&1.57$\pm{0.10}$&82$^{+19}_{-20}$\\
NGC 5548&702042050&14 17 59.5&+25 08 12&0.017&0.017&1.5&2.24$\pm{0.14}$&93$^{+14}_{-12}$\\
NGC 5548&702042060&14 17 59.5&+25 08 12&0.017&0.017&1.5&1.70$\pm{0.08}$&61$^{+21}_{-16}$\\
NGC 5548&702042070&14 17 59.5&+25 08 12&0.017&0.017&1.5&2.03$\pm{0.11}$&89$^{+17}_{-23}$\\
NGC 5548&702042080&14 17 59.5&+25 08 12&0.017&0.017&1.5&3.08$\pm{0.21}$&156$^{+3}_{-2}$\\
NGC 6860&703015010\tablenotemark{5}& 20 08 46.9&--61 06 01&0.015&0.042&1.5&1.42$\pm{0.08}$&106$^{+1}_{-1}$\\
NGC 7213&701029010&22 09 16.3&--47 10 00&0.006&0.020&1.5\tablenotemark{6}&1.44$\pm{0.06}$&89$^{+11}_{-12}$\\
NGC 7314&702015010&22 35 46.2&--26 03 02&0.005&0.015&1 h&1.56$\pm{0.15}$&138$^{+28}_{-23}$\\
NGC 7469&703028010&23 03 15.6&+08 52 26&0.016&0.049&1.5&1.83$\pm{0.06}$&149$^{+13}_{-16}$\\
NGC 985&704042010&02 34 37.8&--08 47 15&0.043&0.029&1\tablenotemark{6}&1.65$\pm{0.17}$&80$^{+30}_{-28}$\\
SWIFT J2127.4+5654&702122010\tablenotemark{5}&21 27 44.9&+56 56 40&0.014&0.787&1&0.90$\pm{0.04}$&159$^{+10}_{-12}$\\
UGC 06728&704029010&11 45 16.0&+79 40 53&0.007&0.045&1.2&1.28$\pm{0.12}$&83$^{+31}_{-29}$\\
\enddata 
\tablenotetext{1}{\citet{Veron-Cetty:2006lr}}
\tablenotetext{2}{The Galactic column density in units of 10$^{22}$ cm$^{-2}$ obtained from the weighted average $N_H$ in the Dickey \& Lockman HI in the Galaxy survey \citep{Dickey:1990uq}}
\tablenotetext{3}{Flux$_{15-50\,keV}$/Flux$_{2-10\,keV}$ with errors calculated from the net count rate errors of the XIS and PIN data}
\tablenotetext{4}{ Total equivalent width of the Fe K$\alpha$ emission line with the errors to 90\% confidence}
\tablenotetext{5}{ Required two Gaussian components to parameterize the Fe K$\alpha$ emission. The EW for these observations were the sum of the narrow and broad base of the Fe K emission}
\tablenotetext{6}{Spectral type as defined in the {\it Swift} BAT 58-month source catalog. Used when no spectral type was available in \citet{Veron-Cetty:2006lr}. }
\label{tab:table}
\end{deluxetable}

\end{document}